\documentclass[letterpaper,prd,aps,superscriptaddress,nofootinbib,showpacs]{revtex4}
\usepackage[latin9]{inputenc}
\usepackage{color}
\usepackage{amsmath}
\usepackage{amssymb}
\usepackage{graphicx}
\usepackage[unicode=true,
 bookmarks=false,
 breaklinks=false,pdfborder={0 0 1},,colorlinks=false]
 {hyperref}

\makeatletter

\@ifundefined{textcolor}{}
{%
 \definecolor{BLACK}{gray}{0}
 \definecolor{WHITE}{gray}{1}
 \definecolor{RED}{rgb}{1,0,0}
 \definecolor{GREEN}{rgb}{0,1,0}
 \definecolor{BLUE}{rgb}{0,0,1}
 \definecolor{CYAN}{cmyk}{1,0,0,0}
 \definecolor{MAGENTA}{cmyk}{0,1,0,0}
 \definecolor{YELLOW}{cmyk}{0,0,1,0}
}

\usepackage{float}
\usepackage{epsfig}

\makeatother

\begin{document}
\begin{flushright}
SMU-HEP-16-06
\end{flushright}
\title{Reconstruction of Monte Carlo replicas from Hessian parton distributions}

\author{Tie-Jiun Hou}
\affiliation{ Department of Physics, Southern Methodist University,\\
 Dallas, TX 75275-0181 U.S.A. }

\author{Jun Gao}
\affiliation{INPAC, Shanghai Key Laboratory for Particle Physics and Cosmology,
Department of Physics and Astronomy, Shanghai Jiao-Tong University, Shanghai 200240, China}
\affiliation{ High Energy Physics Division, Argonne National Laboratory,\\
 Argonne, Illinois, 60439 U.S.A. }

\author{Joey Huston}
\affiliation{ Department of Physics and Astronomy, Michigan State University,\\
 East Lansing, MI 48824 U.S.A. }

\author{Pavel Nadolsky}
\affiliation{ Department of Physics, Southern Methodist University,\\
 Dallas, TX 75275-0181 U.S.A. }

\author{Carl Schmidt}
\affiliation{ Department of Physics and Astronomy, Michigan State University,\\
 East Lansing, MI 48824 U.S.A. }

\author{Daniel Stump}
\affiliation{ Department of Physics and Astronomy, Michigan State University,\\
 East Lansing, MI 48824 U.S.A. }

\author{Bo-Ting Wang}
\affiliation{ Department of Physics, Southern Methodist University,\\
 Dallas, TX 75275-0181 U.S.A. }

\author{Ke Ping Xie}
\affiliation{ Department of Physics, Southern Methodist University,\\
 Dallas, TX 75275-0181 U.S.A. }

\author{Sayipjamal Dulat}
\affiliation{ School of Physics Science and Technology, Xinjiang University,\\
 Urumqi, Xinjiang 830046 China }

\author{Jon Pumplin}
\affiliation{ Department of Physics and Astronomy, Michigan State University,\\
 East Lansing, MI 48824 U.S.A. }

\author{C.-P.\ Yuan}
\affiliation{ Department of Physics and Astronomy, Michigan State University,\\
 East Lansing, MI 48824 U.S.A. }

\begin{abstract}
We explore connections between two common methods for quantifying
the uncertainty in parton distribution functions (PDFs), based on the
Hessian error matrix and Monte-Carlo sampling.
CT14 parton distributions in the Hessian
representation are converted into Monte-Carlo replicas
by a numerical method that reproduces important
properties of CT14 Hessian PDFs: the asymmetry of CT14
uncertainties and positivity of individual parton distributions.
The ensembles of
CT14 Monte-Carlo replicas constructed this way at NNLO and NLO
are suitable for various collider applications, such as cross section
reweighting. Master formulas for computation of asymmetric standard
deviations in the Monte-Carlo representation are derived. A correction
is proposed to address a bias in asymmetric uncertainties
introduced by the Taylor series approximation.
A numerical program is made available for conversion of Hessian PDFs
into Monte-Carlo replicas according to normal, log-normal, and
Watt-Thorne sampling procedures. 
\end{abstract}

\date{January 10, 2017}


\maketitle

\clearpage{}\tableofcontents{}

\section{Introduction}

Modern parton distribution functions (PDFs)  \cite{Dulat:2015mca,Harland-Lang:2014zoa,Ball:2014uwa,Abramowicz:2015mha,Alekhin:2013nda,Accardi:2016qay}
are provided with estimates of uncertainties from multiple origins.
These estimates are essential for understanding of the accuracy of  collider
predictions, both for precision measurements and for new physics searches.  PDFs are
determined from a global statistical analysis of diverse experimental
measurements, in deep-inelastic scattering, in production of vector bosons
and jets, and in other hard-scattering processes. An optimal parametrization
of the PDFs is obtained  by a minimization of the figure-of-merit function ($\chi^{2}$) that quantifies the level of  agreement between experimental data and theoretical predictions. 
The minimization is performed with respect to the PDF parameters of interest,
and with respect to nuisance parameters associated with theory, experiment
and the data analysis procedure. Once the best-fit PDF parametrization
is found, additional parametrizations are constructed for
estimating the total PDF uncertainty. In this paper, we explore the connection between two methods
for quantifying PDF uncertainties, one based on the diagonalization of
the Hessian error matrix \cite{Pumplin:2001ct},
and one on the stochastic (Monte-Carlo)
sampling of parton distributions \cite{Giele:1998gw,Giele:2001mr}.
PDF uncertainties can also be determined by the Lagrange multiplier
\cite{Stump:2001gu} and offset \cite{CooperSarkar:2002yx} methods,
but the PDFs obtained with the Hessian and Monte-Carlo techniques are
the most commonly used.

In any method of  PDF error analysis, the ultimate goal is to provide
information about the probability distribution in the space of PDF parameter values.
This information can be presented in several forms. Much of PDF research
\cite{Dulat:2015mca,Alekhin:2013nda,Abramowicz:2015mha,Harland-Lang:2014zoa,Accardi:2016qay} relies on an analytic $\chi^{2}$ minimization on a class of PDF parametrization
forms, in the same manner as in the CT global analyses.
Since $\chi^{2}(\vec{a})$ can be approximated by a quadratic function
of PDF parameters $\vec{a}$ in the neighborhood of the global minimum,
a boundary of the hyperellipsoid containing
a specified cumulative probability $p$
can be delineated with a relatively small number of error PDFs, corresponding
to the eigenvectors of the Hessian matrix. The PDF uncertainty on
a QCD observable $X$, at a confidence level $p$, is derived via algebraic
``master formulas'' \cite{Pumplin:2001ct,Pumplin:2002vw} from values
$X_{\pm i}$ for $X$ calculated for each eigenvector set $i$. In the 
Gaussian approximation, the  PDF errors on $X$ in the positive and negative
directions are equal. If deviations from the Gaussian behavior are
mild, then the asymmetry between the positive and negative PDF errors can be
estimated in the Hessian approach, too, using master formulas requiring
two error sets per each eigenvector direction \cite{Nadolsky:2001yg}.

An alternative NNPDF Monte-Carlo technique \cite{Forte:2002fg,DelDebbio:2007ee,Ball:2008by,Ball:2012cx,Ball:2014uwa}
provides an ensemble of error PDF sets, or ``replicas'', which sample
the functional forms of parton distributions parametrized by neural
networks, with little bias due to the choice of the parametrization
form. The probability distribution in the hypervolume can in principle
be fully reconstructed, given a sufficiently large ensemble of PDF
replicas.  The prediction for a QCD observable
$X$ is then obtained from the mean and standard deviation of the
values of $X$ for each member of the (large) Monte-Carlo
replica ensemble.

Hessian eigenvector sets can be converted into Monte Carlo replicas
\cite{Watt:2012tq}, and vice versa \cite{Gao:2013bia,Carrazza:2015aoa,Carrazza:2016htc}. Small Hessian eigenvector ensembles are sufficient,
and desirable, for many applications. Monte Carlo calculations based on a
large ensemble of replica PDFs allow the implementation of new experimental
constraints on the PDFs using replica re-weighting \cite{Giele:1998gw,Ball:2010gb,Ball:2011gg}. The full ensemble of Monte Carlo replicas
can be recast into an ensemble
with fewer replicas by unweighting \cite{Ball:2011gg}, by compression
\cite{Carrazza:2015hva}, or by conversion into a Hessian ensemble \cite{Gao:2013bia,Carrazza:2015aoa,Carrazza:2016htc},
while retaining the core statistical information. PDF ensembles from several
groups can be combined by converting the Hessian ensembles into a Monte-Carlo
representation, and then reducing/compressing the combined MC ensemble
into a smaller Hessian or MC ensemble. These techniques are employed
to combine the error PDFs from CT14, MMHT'14, and NNPDF3.0
into smaller PDF4LHC15 ensembles according to the recommendation of 
the PDF4LHC Working Group \cite{Butterworth:2015oua}. 

Conversion of the Hessian eigenvector sets into Monte-Carlo
replica sets is employed in several such applications and influences the
outcomes for the uncertainties and even central predictions. The PDF
conversion must be understood in order to trust this method.
The Watt-Thorne prescription \cite{Watt:2012tq}
provides the simplest realization of the conversion method, under the
approximations of the exactly Gaussian distribution of the PDF parameters
and linear dependence of error PDFs on small variations of the PDF parameters.
It is desirable, however, to develop a formalism that can go beyond
the linear approximation and, given enough Hessian sets, reproduce
non-linear features of the probability distribution such as
asymmetry.

This paper lays out such general formalism. It allows one to
systematically include nonlinear PDF dependence and may
produce different outcomes in some LHC predictions
compared to the Watt-Thorne method, as we will demonstrate.
To apply the general conversion
method, we first construct Monte-Carlo replica ensembles that reproduce both the
\emph{symmetric} and \emph{asymmetric} uncertainties of the CT14 Hessian
sets, and also generate positive-definite replica parametrizations
if desired. The resulting PDF uncertainties with MC replicas are
closer to CT14 asymmetric uncertainties, compared to the Watt-Thorne
method. Reconstruction of the full probability distribution from the Hessian
sets reveals several subtleties, see Sec.~\ref{sec:sec2}B.
In particular, naive Monte-Carlo sampling of the PDFs 
using the Taylor series results in biased asymmetric uncertainty bands that
can be corrected.

It is also instructive to compare statistical properties of
the CT14 MC replica ensemble (obtained by conversion) and NNPDF MC
replica ensemble (obtained using a genetic algorithm). Despite drastic
differences in the two replica generation methods, we demonstrate 
in Sec.~\ref{sec:sec3}B that
the statistical properties of two MC replica ensembles have deep
similarities, and explain why.

In Sec.~\ref{sec:sec2} we compare formulas for the estimation
of Hessian and MC uncertainties and derive prescriptions for the generation
of MC replicas that generalize the Watt-Thorne prescription.
These prescriptions account for asymmetry of PDF errors in the same
manner as in the CT14 Hessian set. At the end of the section, we show
how to construct individual replica sets that reproduce the
positivity requirement imposed on the CT14 Hessian sets.
Sec.~\ref{sec:sec3}
compares PDFs, parton luminosities, predictions for LHC cross sections,
and their uncertainties obtained with the CT14 Hessian and MC PDFs.
In Sec.~\ref{sec:ReplicaInterpretation} we examine statistical properties
of the replica ensemble, and consider the agreement of individual replicas with
the experiments in the global fit. Sec.~\ref{sec:sec4} contains
concluding remarks and information on the availability of the CT14
replica PDFs. A computer program is presented for generation
of MC replicas according to the new prescriptions that optimally reproduce
the Hessian uncertainties and positivity.

\section{Generation of Monte-Carlo replicas from Hessian error PDFs}
\label{sec:sec2}

\subsection{Master formulas for Hessian PDFs}

The CT14 parton distribution functions (PDFs) were generated in a
global analysis of QCD, by fitting theoretical calculations,
at LO, NLO and NNLO,  to experimental
data for a wide variety of high-$Q^{2}$ processes \cite{Dulat:2015mca}.
Once the central (most
probable) combination of PDF parameters is found by minimization of
the log-likelihood function $\chi^2$, special ``error PDF
sets'' are constructed to characterize and
propagate PDF uncertainties. The error PDFs do not specify all
features of the probability distribution in the fit, only a part
needed for estimating
uncertainties in typical applications. In the Hessian method used by CTEQ, the
error PDFs, called ``eigenvector sets'', keep information about the
first and second moments of the probability distribution, sufficient for
estimation of central values, standard deviations, and PDF-driven correlations.
A PDF ensemble based on MC replicas, when obtained directly from the
fit, can in principle
reproduce the primordial probability distribution with better accuracy; in
practice, its size is commonly limited to no more than a thousand
replicas, also enough for reproducing up to the second moments.

Therefore, the error sets primarily quantify the lowest two moments of the
primordial probability; and a Hessian eigenvector set is
often the only published output from the PDF fit. For the applications
utilizing MC replicas, especially when large numbers of replicas are to
be generated on the fly, one may wish to convert the Hessian PDF ensemble
into an MC replica one, according to the procedure that will be
laid out. [A computer code implementing this procedure is available
  for downloading.] The converted MC replicas must ideally preserve
the statistical
features of the Hessian ensemble. They cannot provide more
information about the primordial higher moments than what
is available in the Hessian ensemble. 

Let us focus for a moment on the construction and properties
of the CT Hessian sets.
The PDFs $f_{n}(x,Q)$ are parametrized at $Q=Q_{0}=1.3\ {\rm GeV}$
by a functional form $F_{n}(x,\vec{a})$ which depends on a set of $D$
adjustable parameters, $\vec{a}=\{a_{1},a_{2},a_{3},...,a_{D}\}$. A
test statistic $\chi^{2}(\vec{a})$ is defined as in \cite{Lai:2010vv,Gao:2013xoa}
to measure the difference between theory and data. Its explicit
definition is reproduced in the appendix. The minimum of
$\chi^{2}(\vec{a})$ defines the central fit, and the variation of $\chi^{2}$
in a neighborhood of the minimum provides the uncertainty in the fit.
Thus, the approximation in the vicinity of the minimum,
\begin{equation}
\chi^{2}(\vec{a})\approx\chi^{2}\left(\vec{0}\right)+\frac{1}{2}\sum_{i,j=1}^{D}\left(\frac{\partial^{2}\chi^{2}}{\partial{a_{i}}\partial{a_{j}}}\right)_{\vec{a}=\vec{0}}a_{i}a_{j}=\chi_{0}^{2}+\sum_{i,j=1}^{D}H_{ij}a_{i}a_{j},
\end{equation}
has two parts. First, the central combination of the parameters, which corresponds,
without loss of generality, to all zero
$a_{i}$ values ($\vec{a}=\{0,\,0,\,...0\}\equiv\vec{0})$, gives the ``best''
fit between theory and data, for which $\chi^{2}$ is minimized: $\chi^{2}(\vec{0})=\mbox{min }\chi^{2}\equiv\chi_{{\rm 0}}^{2}.$
Second, the approximately quadratic behavior of $\chi^{2}$ very close
to the minimum allows one to define independent directions in $\vec{a}$
space, corresponding to eigenvectors $\widehat{z}_{i}$ of the Hessian
matrix $H_{ij}.$ The eigenvectors are used to determine the uncertainty
on PDFs $f$ and on any quantities $X(f)$ that depend
on them, by computing variations of $\chi^{2}$ along each independent
direction \cite{Pumplin:2001ct}.

\begin{figure}
\includegraphics[width=0.9\columnwidth]{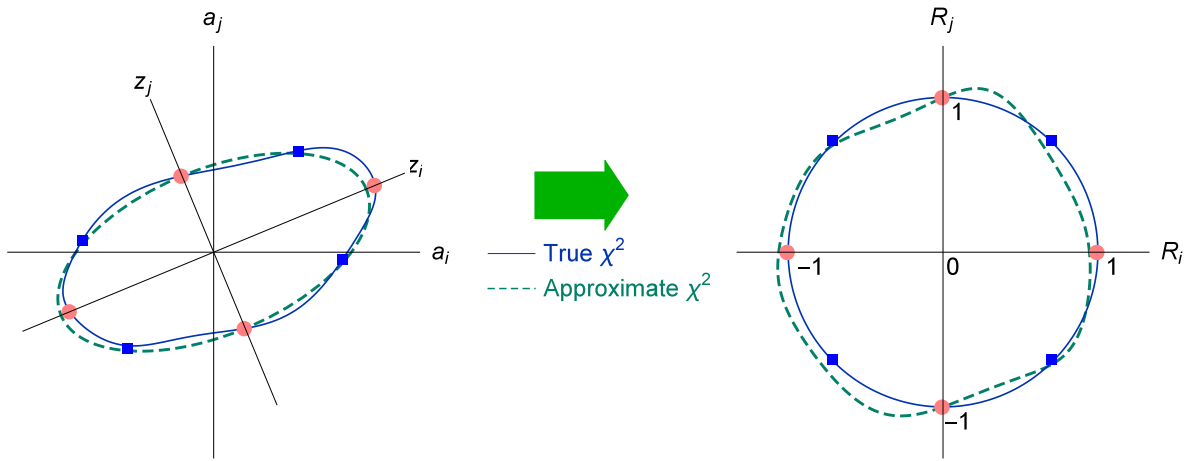}

\caption{Contours of constant $\chi^{2}$ under the scaling transformation
of PDF parameters. \label{fig:Gaussian-contours}}
\end{figure}

In this scheme, the independent directions $\widehat{z}_{i}$ are determined
by relying on the quadratic approximation for $\chi^{2}$;  at the
same time, extreme displacements $\bar{z}_{\pm i}$ along each eigenvector
direction are found from variations of the true (not perfectly quadratic)
$\chi^{2}.$ This is illustrated in the left inset of
Fig.~\ref{fig:Gaussian-contours},
showing contours of constant values of the exact and approximate $\chi^{2}$ for
a certain pair of parameters, $\{z_{i},z_{j}\}$.
The minimum of the exact $\chi^{2}$
is reached at the axes origin; the solid and dashed contours are drawn
for constant increases of $\Delta\chi_{i}^{2}$ with respect to the minima
of the exact $\chi^{2}$ and its quadratic approximation. The $\widehat{z}$
directions are along the axes of the approximate $\chi^{2}$ ellipsoid.
The extreme displacements $\bar{z}_{\pm i}$ are found by demanding the
increase $T^{2}$ in the true, not approximate, $\chi^{2}$. They
correspond to the (filled) red circles at the intersections
of the $z$ axes and the irregular contours of the true $\chi^{2}$
in the left-hand side of Fig.~\ref{fig:Gaussian-contours}. Note that
both the functional form of the true $\chi^2$ and the value of $T^2$
depend on the context of the analysis. In CT fits
\cite{Lai:2010vv, Gao:2013xoa},
the true $\chi^2$ that determines the uncertainties
consists of two tiers of conditions,
quantifying the global agreement with the experiments, and deviations
from individual experiments, cf. the appendix. The parameter $T$ is set to 10 at 90\%
confidence level (c.l.).  
The complex nature of true $\chi^2$ implies that, most generally,
$\bar{z}_{+i}\neq \bar{z}_{-i}$; still, its overall features are captured well by the
approximate quadratic $\chi^2$. 

This introduces one source of asymmetry in PDF errors, due to
the cubic and higher sign-odd powers of $z$ in the full $\chi^2(z)$. The impact
of this source is reduced by rescaling
$z_{_{i}}\rightarrow R_{i}(z_{i})$ of the parameters to satisfy $R_{i}(\vec{0})=0,$
$R_{i}(\bar{z}_{\pm i})\equiv R_{\pm i}=\pm 1.$
Keep in mind that disagreements of
the primordial $z$-dependent probability with the Gaussian behavior
are weak to start with, and that the Hessian sets specify the
probability (i.e., $\chi^2$)
only at three points per an eigenvector direction. {\it Aside from the
  above conditions at the three points, we will not
  need to know the rescaling function $z_i(R_i)$
  in the rest of the discussion.}\footnote{An example of the
  rescaling function for one eigenvector direction is a steplike one,
  $z_i(R_i)=\left| R_i/2 \right| \left[\bar{z}_{+i}+\bar{z}_{-i}+(\bar{z}_{+i}-\bar{z}_{-i})\mbox{tanh}(R_i/\epsilon)\right]$.
  The parameter $\epsilon$ controls the smoothness of the step
transition, it is at least as small as
$(\bar{z}_{+i}+\bar{z}_{-i})/(\bar{z}_{+i}-\bar{z}_{-i})$.
Far enough from $z_i=R_i=0$, the
step function reduces to $z_i=|R_i|\bar{z}_{+i}$ and $z_i=|R_i|\bar{z}_{-i}$, for
positive and negative $z_i$, respectively.}
The probability
distribution in terms of $\vec{R}$ is even closer to the Gaussian one:
\begin{equation}
{\cal P}\left(\vec{R}\right)\approx\frac{1}{(2\pi)^{D/2}}\exp\left(-\frac{1}{2}\sum_{i=1}^{D}R_{i}^{2}\right).\label{PR}
\end{equation}
The transformation is illustrated by the right-hand side of
Fig.~\ref{fig:Gaussian-contours}.
This approximation is known from experience to hold well
in CT fits within the 90\% c.l. region around
the best fit, even though large $\left|R_{i}\right|$ (greater than 1.5) may
be excluded by tier-2 penalties on $\chi^{2}$ that enforce agreement
with the individual fitted experiments \cite{Lai:2010vv,Gao:2013xoa}.

Next, we consider a function $X(\vec{R})$ of the parameters $R_i$,
such as the PDF or even the QCD cross section.
This function may have asymmetric uncertainties, regardless of
symmetry of ${\cal P}\left(\vec{R}\right)$.
Expanding $X(\vec{R})$ near the global $\chi^{2}$ minimum
in the Taylor series up to the second derivatives, we have
\begin{equation}
X(\vec{R})=X_{0}+\sum_{i=1}^{D}\frac{\partial X}{\partial R_{i}}R_{i}+\frac{1}{2}\sum_{i,j=1}^{D}\frac{\partial^{2}X}{\partial R_{i}\partial R_{j}}R_{i}R_{j}+...\,.\label{Taylor}
\end{equation}
The derivatives can be estimated by finite differences,
\begin{eqnarray}
\frac{\partial X}{\partial R_{i}} & \approx & \frac{X_{+i}-X_{-i}}{2},\\
\frac{\partial^{2}X}{\partial R_{i}^{2}} & \approx & X_{+i}+X_{-i}-2X_{0},\\
\frac{\partial^{2}X}{\partial R_{i}\partial R_{j}} & \approx & \frac{X_{+i,+j}+X_{-i,-j}-X_{+i,-j}-X_{-i,+j}}{2}.
\label{dnXdRn}
\end{eqnarray}
In these equations, the $X$ values are taken at the global minimum,
\begin{eqnarray}
X_{0} & \equiv & X\left(\vec{0}\right);
\end{eqnarray}
at the red circles on the axes in Fig.~\ref{fig:Gaussian-contours},
\begin{eqnarray}
X_{\pm i} & \equiv & X\left(0,0,...,R_{i}=\pm1,...,0\right);
\end{eqnarray}
and at the blue squares on the diagonals,
\begin{eqnarray}
X_{\pm i,\pm j} & \equiv & X\left(0,...,R_{i}=\pm\frac{1}{\sqrt{2}},...,R_{j}=\pm\frac{1}{\sqrt{2}},...,0\right).
\end{eqnarray}
We see that the $2D$ Hessian eigenvector sets corresponding to $R_{i}=\pm1$
are sufficient for computing the first and diagonal second derivatives,
$\partial X/\partial R_{i}$ and $\partial^{2}X/\partial R_{i}^{2}$.
The mixed second derivatives would require $2D(D-1)$ additional error sets
for $\{R_{i},R_{j}\}=\{\pm2^{-1/2},\pm2^{-1/2}\}$. Those off-diagonal error sets are not provided in the conventional PDF analyses.

The Taylor expansion of $X\left(\vec{R}\right)$ can be used to construct
various master formulas for the estimation of Hessian PDF uncertainties.
Keeping only linear terms on the right-hand side of Eq.~(\ref{Taylor}),
one obtains
\begin{equation}
X(\vec{R})=X_{0}+\sum_{i=1}^{D}\frac{X_{+i}-X_{-i}}{2}R_{i}\,.\label{Taylor1}
\end{equation}
If $R_{i}=\pm1$ are reached at the boundary of the 68\% c.l. region
(and the Hessian eigenvector sets are determined according to this
convention), the maximal variation of $X$ within the unit hypersphere
of $\vec{R},$
\begin{equation}
\delta_{68}^{H}X=\left|\nabla X\right|=\frac{1}{2}\sqrt{\sum_{i=1}^{D}\left[X_{+i}-X_{-i}\right]^{2}} ,\label{dH68X}
\end{equation}
yields the Hessian symmetric PDF uncertainty at 68\% c.l., denoted
by $\delta_{68}^{H}X$ \cite{Pumplin:2001ct}. However,
the published CT14 parametrizations correspond to the 90\% c.l. Using
the same master formula, we find the 90\% c.l. uncertainty from the
CT14 sets, satisfying $\delta_{90}^{H}X\approx1.65\,\delta_{68}^{H}X$.

If the {\em diagonal} second-order derivatives are also included, Eqs.~(\ref{Taylor1})
and (\ref{dH68X}) are modified to
\begin{equation}
X(\vec{R})=X_{0}+\sum_{i=1}^{D}\left(\frac{X_{+i}-X_{-i}}{2}R_{i}+\frac{X_{+i}+X_{-i}-2X_{0}}{2}R_{i}^{2}\right).\label{Taylor2}
\end{equation}
Also, asymmetric uncertainties on $X$ can be estimated by \cite{Nadolsky:2001yg}
\begin{eqnarray}
\delta_{68}^{H>}X & = & \sqrt{\sum_{i=1}^{D}\left[\mbox{max}\left(X_{+i}-X_{0},\, X_{-i}-X_{0},\,0\right)\right]^{2}} ,\label{dH68Xp}\\
\delta_{68}^{H<}X & = & - \sqrt{\sum_{i=1}^{D}\left[\mbox{max}\left(X_{0}-X_{+i},\, X_{0}-X_{-i},\,0\right)\right]^{2}} .\label{dH68Xm}
\end{eqnarray}
The absolute values of the asymmetric uncertainties
are generally not the same for the positive ($>$) and
negative ($<$) displacements \cite{Nadolsky:2001yg}. For example, $X_{+i}-X_{0}$
and $X_{-i}-X_{0}$ must have opposite signs if $X$ varies linearly with
$R_i$, but they can have the same sign if the nonlinearity of $X$ is large and hence
result in a zero contribution to either $\delta_{68}^{H>}X$ or $\delta_{68}^{H<}X$.
In the CT convention, low-number eigenvectors, corresponding
to the best-determined directions, tend to be more symmetric
(with opposite-sign variations), while high-number eigenvectors can be strongly
asymmetric.

\subsection{Probability density distribution for $X$}
\label{sec:PX}
We will now formulate a set of requirements for Monte-Carlo replica
generation that will be applied in the next subsection to devise our replica
generation method. A reader interested in the
specific formulas for replica generation may skip 
this subsection and proceed to the next. 

The Monte-Carlo replicas quantify the full probability density distribution
$\mathcal{P}(X)$ for a QCD observable $X$, going beyond the intervals
of fixed probability available with the Hessian PDFs. A priori, 
$\mathcal{P}(X(\vec R))\neq \mathcal{P}(\vec R)$.
The probability $\mathcal{P}\left(X(\vec{R})\right)$
might be extracted \emph{directly} from the fit by using a Lagrange
multiplier scan or stochastic sampling of the \emph{exact} log-likelihood
$\chi^{2}$. But, more often than not, the outcomes of the fit are
initially stored in the form of the Hessian PDFs $f_{a}(x,Q;\vec{R})$. The complete functional forms for the PDFs may be unavailable or difficult to sample.
Then, one relies on constructing the Monte-Carlo replicas from the Taylor
series for the Hessian PDF values provided in $\{x,Q\}$ space, and without invoking the exact parametrization form.

In addition to the central prediction $X_{0}$, the
Hessian PDFs specify two values of $X$ per each eigenvector direction,
that is, $X_{\pm i}$. This information is generally
insufficient for reconstructing the true $\mathcal{P}(X)$ by Monte-Carlo
sampling. Even the confidence of probability intervals
for $\mathcal{P}(X)$ may remain undetermined! 

Recall that, by their construction and using
the parameter distribution $\mathcal{P}(\vec{R})$,
the 68\% c.l. Hessian values $\{X_{-i},\,X_{0},\,X_{+i}\}$
correspond to the cumulative probabilities of about 16\%, 50\%, and
84\%. When one
naively samples $X(\vec{R})$ assuming a multi-Gaussian distribution
for $\vec{R},$ one is not guaranteed to obtain the same cumulative
probabilities for $\{X_{-i},\,X_{0},\,X_{+i}\}$ from the distribution
$\mathcal{P}(X)$ of Monte-Carlo replicas, because of \emph{probability}
\emph{folding}. When folding occurs, the 68\% probability interval
of the Monte-Carlo replica set distributed according to $\mathcal{P}(X)$
deviates from the 68\% c.l. Hessian uncertainty interval found using
$\mathcal{P}(\vec{R})$. In the Gaussian case, the central combination
$\vec{R}=\vec{0}$ of the PDF parameters, associated with the central
CT14 PDF, is located at the mode, median, and mean of $\mathcal{P}(\vec{R})$,
as those coincide. On the other hand, the central CT14 prediction for $X$
does not automatically correspond to either the mode, median, or mean
of $\mathcal{P}(X),$ which can be different.

Indeed, when $X(\vec{R})$ is not monotonic, several combinations $\left(\vec{R}(X)\right)_k$
may produce the same value of $X.$ Replicas of such $X$ obey a
folded probability $\mathcal{P}(X)$, meaning
that the cumulative probability for getting $X(\vec{R})$ is not equal
to that for the corresponding $\vec{R}$. In the one-dimensional case, a simple
example is $X=\left|R\right|$ with a normally distributed $R$ on
$[-\infty,+\infty]$. Then $X$ obeys a half-normal distribution on
$X\in[0,+\infty]$; $X_{0}=0$ corresponds to the cumulative probabilities
of 50\% and 0\% according to $\mathcal{P}(R)$ and $\mathcal{P}(X(R))$,
respectively. 

If $X(R)$ depends on one eigenvector parameter $R$ and is continuous and monotonic
in $K$ intervals $[r_{0},r_{1}]$,..., $[r_{k-1},r_k]$, then $X(R)$ is invertible
in each interval: $R=R_{k}(X)$ for $r_{k-1}\leq R\leq r_{k}$.
We indicate that $x$ lies within $[a,b]$ using a notation
$\Theta\left(x\in[a,b]\right),$ with $\Theta(L)$ equal to $1$
or 0 if the statement $L$ is true or false, respectively. 

The function $X(R)$ is generally not normally distributed, even when
$R$ is. That is, while 
\begin{equation}
\mathcal{P}(R)=\frac{1}{(2\pi)^{1/2}}e^{-R^{2}/2},
\end{equation}
the probability density distribution $\mathcal{P}(X)$ does not need
to be Gaussian. It satisfies
\begin{equation}
\mathcal{P}(X)=\sum_{k=1}^{K}\Theta\left(R\in[r_{k-1},r_{k}]\right)\,\Theta\left(X\in[X(r_{k-1}),X(r_{k})]\right)\mathcal{\,P}\left(R_{k}(X)\right)\,\frac{dR_{k}(X)}{dX},\label{PXPR}
\end{equation}
as follows from the normalization conditions for the probability densities,
\begin{equation}
\int\mathcal{P}(R)dR=\int\mathcal{P}(X)dX=1.
\end{equation}

Going back to the truncated Taylor series with a non-negligible $\partial^{2}X/\partial R^{2}$,
\begin{equation}
X(R)=X_{0}+\frac{\partial X}{\partial R}R+\frac{1}{2}\frac{\partial^{2}X}{\partial R^{2}}R^{2}+...,\label{XR1}
\end{equation}
where the derivatives are estimated by the finite differences
from $X_0$  and $X_\pm$, we could solve for $R(X)$ and find up
to \emph{two} solutions for $R$ per each $X$ value. {[}$\mathcal{P}(X)$
is then given by Eq.~(\ref{PXPR}) with $K=2$.{]} Such truncated
Taylor approximation is clearly not monotonic. We expect it to deviate
from the true $X(R)$ when the displacements $R$ are large, and to
produce a folded probability distribution $\mathcal{P}(X)$ with a
displaced median value and confidence intervals, as compared to the
true $\mathcal{P}(X)$. We may even end up with a problematic arrangement when the central CT14 Hessian prediction is outside of the nominal 68\% probability interval of the replicas. These observations continue to hold when $\vec{R}$ has $D$ components.

On the other hand, in a variety of important applications, our group
observed that the Lagrange
multiplier and Hessian methods render similar probability intervals.
Most recently, the two methods were shown to predict compatible
68\% and 90\% confidence intervals for the common LHC observables,
such as jet, Higgs, and $t\bar{t}$ production
cross sections \cite{Dulat:2015mca, Belyaev:2005nu, Dulat:2013kqa, Schmidt:2014gda, Pumplin:2009nk}.
This indicates that
the impact of the second derivatives is mild in constrained regions,
credible estimates for
$\mathcal{P}(X)$ can be found via the sampling of the Hessian PDFs
after correcting for the Taylor series' imperfections. At intermediate
$x,$ where the nonlinearities in the PDF uncertainties are weak,
the Taylor expansion performs well. The discrepancies are more pronounced
at small and large Bjorken $x$ with fewer experimental constraints.
Nonlinearities and folding effects are small for well-constrained
eigenvector parameters $R_{i}$ and can be substantial for poorly
constrained ones. 

With these observations in mind, we design a Monte-Carlo sampling procedure 
for the PDFs using the Taylor
series in the case of mild nonlinearities. First, rather
than always sampling the PDFs $f_{a}(x,Q;\,\vec{R})$, we may
sample a function of $X$ of $f_{a}(x,Q;\,\vec{R})$ that reduces the
impact of $\partial^{2}X/\partial R_{i}^{2}$. For example, at moderate
$x,$ one could just sample the PDFs directly, i.e., set $X=f_{a}(x,Q;\,\vec{R})$
with normally distributed $\vec{R}$. At large $x,$ where $f_{a}(x,Q)\sim(1-x)^{R}$,
Gaussian sampling of $X\equiv\ln\left(f_{a}(x,Q)\right)\sim R\cdot const$
may be more desirable for suppressing the nonlinearities; we will
show that the log-normal sampling leads to more physical behavior. 
We explore this possibility when generating positive-definite PDF
replicas in Sec.~\ref{sec:MCPDFs}.

Second, we need to assign cumulative probabilities to the input $X$ values.  Within the accuracy of the Hessian approximation, we can 
make use of the knowledge that the Hessian central value
and Hessian uncertainties are close to the median and
68\% central probability intervals for the PDFs, respectively. This
``obvious'' conclusion from the Lagrange Multiplier studies
can be violated
by naive sampling, hence we enforce it as a separate requirement. Furthermore,
for mildly asymmetric populations, the median, with the cumulative
probability of 0.5, tends to lie close to the mean given by the standard
formula \cite{KenneyKeeping}.
In the various sampling scenarios that we had
tried, we confirmed that the median of a PDF replica distribution differed little from the mean, except for in the extreme regions.
It suffices to assume
that the central Hessian PDF coincides with the \emph{mean} of the
PDF replica ensemble, easily computable from the replicas' values and close to the median value.
When the replica ensemble is first generated, the median/mean at each
$x$ and $Q$ may disagree from the central CT14 PDF as a consequence
of the probability folding caused by the Taylor expansion,
as well as because of residual
fluctuations. We correct for this mismatch by a uniform shift in all PDF
replica values, as explicated in the next section. 

In Sec.~\ref{sec:sec3}, we show that this sampling procedure reproduces 
the central CT14 Hessian prediction (coinciding with the mean of the
MC replicas by construction), and the 68\% probability intervals of
the MC replicas approximate both the Hessian symmetric and asymmetric
uncertainties. 

\subsection{Generation of Monte-Carlo replicas \label{sec:GenMC}}

We will now present the formulas to generate the 
Monte-Carlo PDF replicas. We produce $N_{rep}$
sets of the PDF parameters, $\vec{R}^{(k)}\equiv\left\{ R_{1}^{(k)},...,R_{D}^{(k)}\right\} $,
which are randomly distributed according to the probability density
${\cal P}\left(\vec{R}\right)$.  We will again rely on
the experience that ${\cal P}\left(\vec{R}\right)$ is
typically close to the standard normal distribution, as in Eq.~(\ref{PR}),
with the mean (or, central) values $\vec{R}=\vec{0}$ corresponding to
the best-fit PDF set, and the standard deviations that are
found from the Hessian
PDFs. Therefore the replicas $\vec{R}^{(k)}$ will be sampled from the standard
normal distribution in Eq.~(\ref{PR}).

The Monte-Carlo replica $X^{(k)}$ of a QCD observable, $X$, will be
constructed as
\begin{equation}
X^{(k)}=X_{0}+d^{(k)}-\Delta ,\label{Xk}
\end{equation}
in terms of the value $X_{0}$ for the central Hessian set, a random
shift $d^{(k)}\equiv d\left(\vec{R}^{(k)}\right)$, and a constant shift
$\Delta$ applied to all replicas ($k=1,...,N_{rep})$. Examples of
$X(\vec{R})$ include the PDF, parton luminosity, cross section, or the logarithm of PDF discussed below. The $\Delta$
parameter can be set to zero, as in the original
Watt-Thorne method. But as was argued in Sec.~\ref{sec:PX}, we
find it helpful to allow a shift of all replica values by a constant
amount $\Delta$.

When $X$ is a PDF, we
expect that $\langle X(\vec R)\rangle \approx X(\langle \vec R\rangle) = X_0$
within the accuracy of the Hessian approximation. [This relation does
  not hold for an arbitrary $X$.] 
Monte-Carlo sampling of the Taylor series may disagree
with this condition. The Taylor expansion 
may lead to systematic biases, as discussed
above. That is, $X_0$ may differ from the mean $\langle X\rangle$ over
the replica ensemble, 
\begin{equation}
\langle X\rangle\equiv\frac{1}{N_{rep}}\sum_{k=1}^{N_{rep}}X^{(k)},\label{Xav}
\end{equation}
even when $N_{rep}$ is large. The mean $\langle X\rangle$ also fluctuates
around $X_{0}$, as the number of replicas changes, but these fluctuations
are small when $N_{rep}$ is above a few hundred. If our goal is to
reproduce the Hessian central
value (``truth'') and Hessian uncertainties as closely as possible, we may
correct such small offsets of the mean, after all $d^{(k)}$
are computed, by shifting all replicas by a constant amount
\begin{equation}
\Delta\equiv\langle d\rangle=\frac{1}{N_{rep}}\sum_{k=1}^{N_{rep}}d^{(k)}.\label{Delta}
\end{equation}
We then get $\langle X\rangle=X_{0}$, up to a small numerical
uncertainty, when the replicas are generated according to Eqs.~(\ref{Xk})
and (\ref{Delta}). This choice will be adopted as the default. 
The random displacements $d^{(k)}$ are found using
Eq.~(\ref{Taylor}), where $R_{i}=R_{i}^{(k)}$
for the $k$-th replica.
For the PDFs, $X(\vec{R})=f_{a}(x,Q_0; \vec{R})$,
this prescription preserves the usual sum rules
 obeyed by the PDFs, since each replica is a linear combination of
 the Hessian eigenvector sets, which satisfy the sum rules individually.

\textbf{Symmetric MC replicas.} If only the first derivatives are retained
in the Taylor series, then we have
\begin{equation}
d^{(k)}=\sum_{i=1}^{D}\frac{X_{+i}-X_{-i}}{2}R_{i}^{(k)}.\label{dXkSym}
\end{equation}
On average, $X^{(k)}$ are symmetrically distributed in the positive
and negative directions. The mean of the replicas, remains close
to the central Hessian value $X_{0}:$ $\langle X\rangle=X_{0}$ with
good accuracy, and the global shift $\Delta$ can be neglected.
The MC uncertainty on $X$
is then quantified by a standard deviation,
\begin{equation}
\delta^{MC}X=\sqrt{\langle\left(X-\langle X\rangle\right)^{2}\rangle} .\label{dMCX}
\end{equation}
It is expected to approach the 68\% c.l. symmetric Hessian uncertainty
$\delta_{68}^{H}X$ in Eq.~(\ref{dH68X}),
in the $N_{rep}\rightarrow\infty$ limit.

\textbf{Asymmetric MC replicas. }In analogy to Eq.~(\ref{Taylor2})
for the Hessian uncertainties, the {\em diagonal} second derivatives $\partial^{2}X/\partial R_{i}^{2}$
can be included as
\begin{equation}
d^{(k)}=\sum_{i=1}^{D}\left(\frac{X_{+i}-X_{-i}}{2}R_{i}^{(k)}+\frac{X_{+i}+X_{-i}-2X_{0}}{2}\left(R_{i}^{(k)}\right)^{2}\right).\label{dXkAsym}
\end{equation}
In this case Eq.~(\ref{Xk}) must include the shift term
$\Delta=\langle d\rangle$ to satisfy $\langle X\rangle = X_0$. Now the
{\em asymmetric} error estimators are given by the standard deviations that
include only positive (negative) displacements from the mean value:
\begin{eqnarray}
\delta^{MC>}X & = & +\sqrt{\langle\left(X-\langle X\rangle\right)^{2}\rangle_{X>\langle X\rangle}} ,\label{dMCXp}\\
\delta^{MC<}X & = & -\sqrt{\langle\left(X-\langle X\rangle\right)^{2}\rangle_{X<\langle X\rangle}} .\label{dMCXm}
\end{eqnarray}
These are the MC counterparts of the asymmetric Hessian uncertainties, $\delta_{68}^{H\gtrless},$
provided by Eqs.~(\ref{dH68Xp}) and (\ref{dH68Xm}).

Compared to the previous work on the MC replicas, our prescription thus includes several
new features: the generation of MC replicas using asymmetric displacements
(\ref{dXkAsym}) derived from the Taylor series, a constant shift
$\Delta$ to ensure agreement between the central Hessian set and the mean of the
replicas, and asymmetric standard deviations $\delta^{MC\gtrless}X$.
Watt and Thorne \cite{Watt:2012tq} provided the same formulas
(\ref{dXkSym}, \ref{dMCX})
for generating the symmetric MC replicas. Their prescription for generating
MC replicas with asymmetric displacements is different from the Taylor-series
displacements in (\ref{dXkAsym}) and is given by
\begin{equation}
d^{(k)}=\sum_{i=1}^{D}\left(\left(X_{+i}-X_{0}\right)\,\Theta\left[R_{i}^{(k)} > 0\right]+\left(X_{-i}-X_{0}\right)\,\Theta\left[R_{i}^{(k)} < 0 \right]\right)\left|R_{i}^{(k)}\right|,\label{XkAsymWT}
\end{equation}
with $\Delta=0.$

We advocate using the asymmetric standard deviations (\ref{dMCXp}) and (\ref{dMCXm})
as estimators for 68\% c.l. PDF uncertainties, as they are simple, numerically close to the Hessian asymmetric estimators, cf. the next section, 
and converge rapidly with the number of replicas.\footnote{An alternative estimator, given by the 68\% probability interval
centered on the mean value, requires sorting the $X^{(k)}$ values
and is more susceptible to random fluctuations. Another estimator,
given by the maximal displacements of $X$ in the 68\% probability hypersphere for $\vec R$, has slow convergence.}
In terms of the relative significance, using $\delta^{MC\lessgtr}X$
with $\Delta$ shifts, together with the asymmetric standard deviations,
is most important for reproducing the asymmetric
Hessian uncertainties. As a secondary effect, mild numerical differences were also noticed between the Taylor-series formula (\ref{dXkAsym}) and Watt-Thorne formula (\ref{XkAsymWT}) for displacements $d^{(k)}$.

\subsection{Monte-Carlo replicas for CT14 parton distributions \label{sec:MCPDFs}}

The formulas derived in Sec.~\ref{sec:GenMC}
are applicable for generating MC replicas for the PDFs themselves,
with an extra modification. We perform 
a truncated Taylor series expansion of $f_a (x,Q; \vec{R})$ at every
$x$ point, assuming 
small deviations from the central value $X_{0}$.
In unconstrained $x$ regions, the linear expansion
is not sufficient -- in fact, it may lead to unphysical solutions
such as negative cross sections. Parametrizations for Hessian distributions
from the CT family satisfy positivity constraints, $f_{a}(x,Q)\geq0$, 
and render non-negative physical cross sections. To realize
the positivity of each MC replica, and eventually to better
approximate the non-linear probability distribution,
we construct the replicas by Gaussian sampling
of $\ln\left[f_{a}(x,Q)\right]$, rather than by sampling $f_{a}(x,Q)$ directly.

The final distribution of CT14 MC replicas includes two families at
NNLO in the QCD coupling strength, with 1000 replica sets each, as
well as two counterpart families at NLO. Replicas $f_{a}^{(k)}(x,Q)\equiv f^{(k)}$
of the first type (linear MC, or MC1) are generated assuming normally distributed linear
displacements of PDFs, corresponding to Eqs.~(\ref{Xk}), (\ref{dXkAsym})-(\ref{dMCXm}),
with $X=f.$ Replicas of this type are very similar in spirit to Watt-Thorne
replicas; in particular, they may violate positivity.

Replicas of the second type (log MC, or MC2) are generated by sampling of a Gaussian
distribution for $X=\ln\left[f\right],$ whose variance is estimated
as $\left(\delta^{MC}X\right)^{2}$, and the mean value is set to
$\mu=\ln\left[f_{0}\right]-\Delta-\left(\delta^{MC}X\right)^{2}/2$,
where $\Delta=\langle d\rangle,$ as above. For the PDFs $f=\mbox{exp}\left[X\right]$, we obtain a log-normal distribution with $\langle f\rangle\approx f_{0}$.
When the PDF error is small, the variance of the PDFs, $\mbox{var}\left[f\right]=f_{0}^{2}\left[e^{\left(\delta^{MC}X\right)^{2}}-1\right],$
becomes equal to the squared Hessian error, $\mbox{var}\left[f\right]\approx\left(\delta_{68}^{H}f\right)^{2}$. Recall that,  if $X$ obeys a normal
distribution with the mean $\mu$ and variance $\sigma$, then the
mean and variance of $f=\exp\left[X\right]$ are $\langle f\rangle=\exp\left[\mu+\sigma^{2}/2\right]$
and $\mbox{var}\left[f\right]=e^{\mu+2\sigma^{2}}\left(e^{\sigma^{2}}-1\right)$.
Consequently the ``log MC'' replicas reproduce the Hessian PDF uncertainty
in well-constrained $x$ regions and stay non-negative in the poorly
constrained regions.

\begin{figure}[tb]
\includegraphics[width=0.46\textwidth]{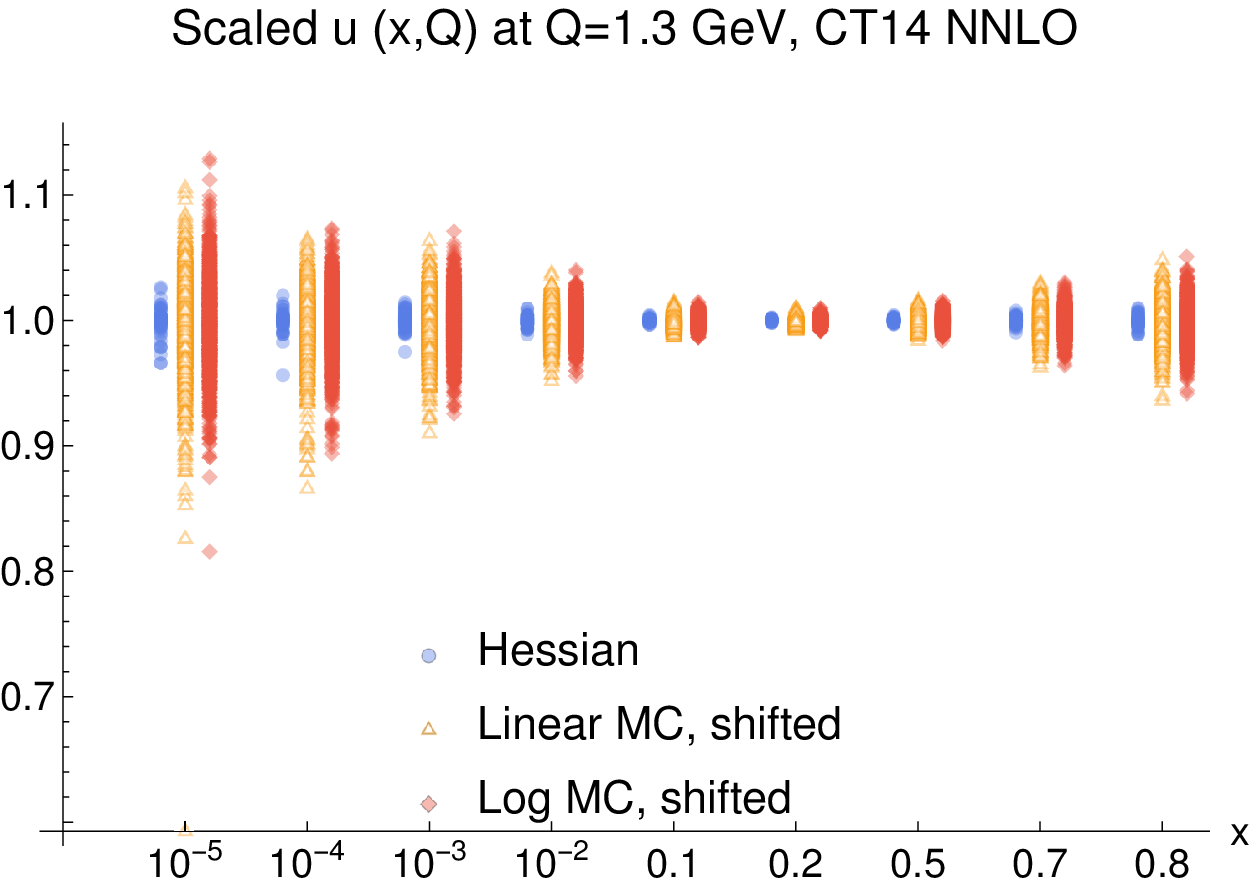}\includegraphics[width=0.46\textwidth]{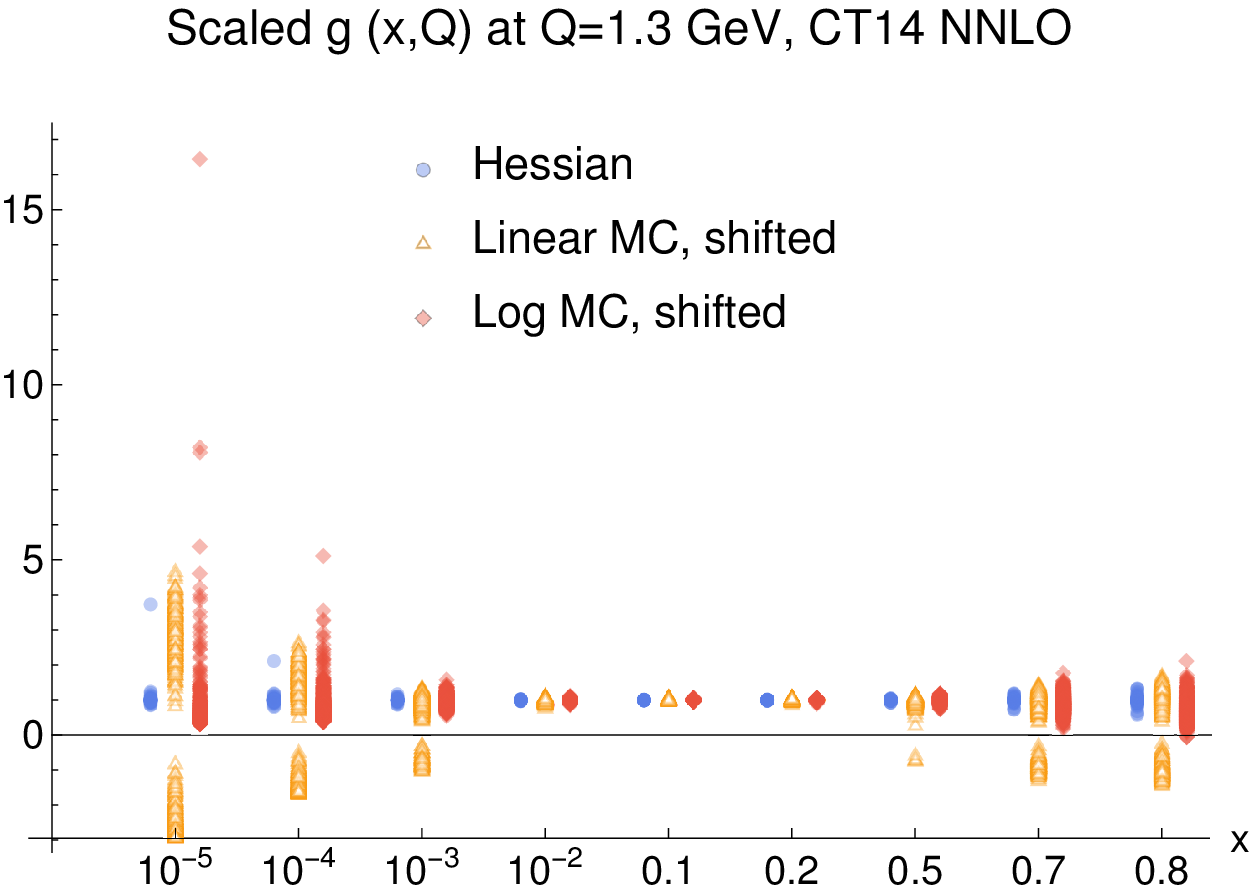}

\includegraphics[width=0.46\textwidth]{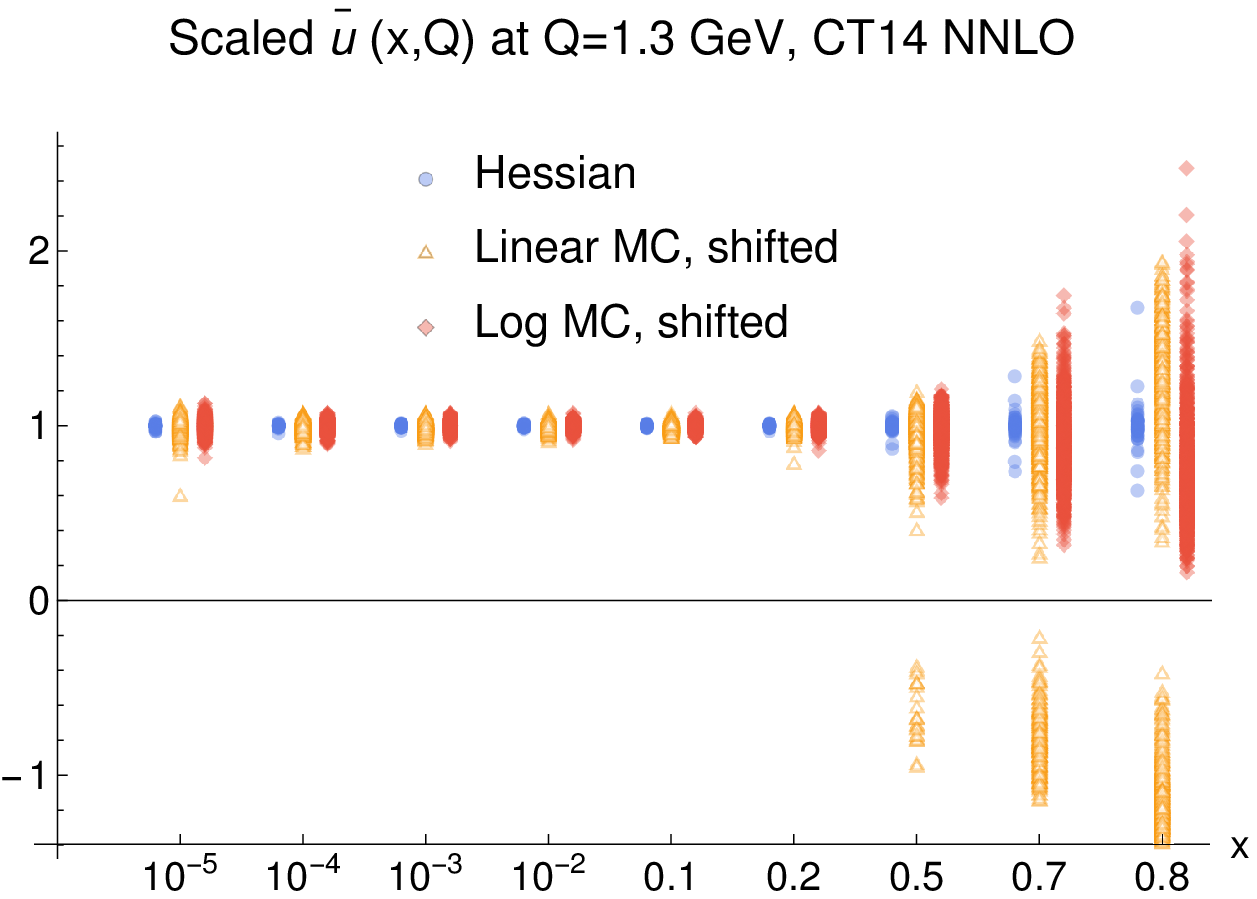}\includegraphics[width=0.46\textwidth]{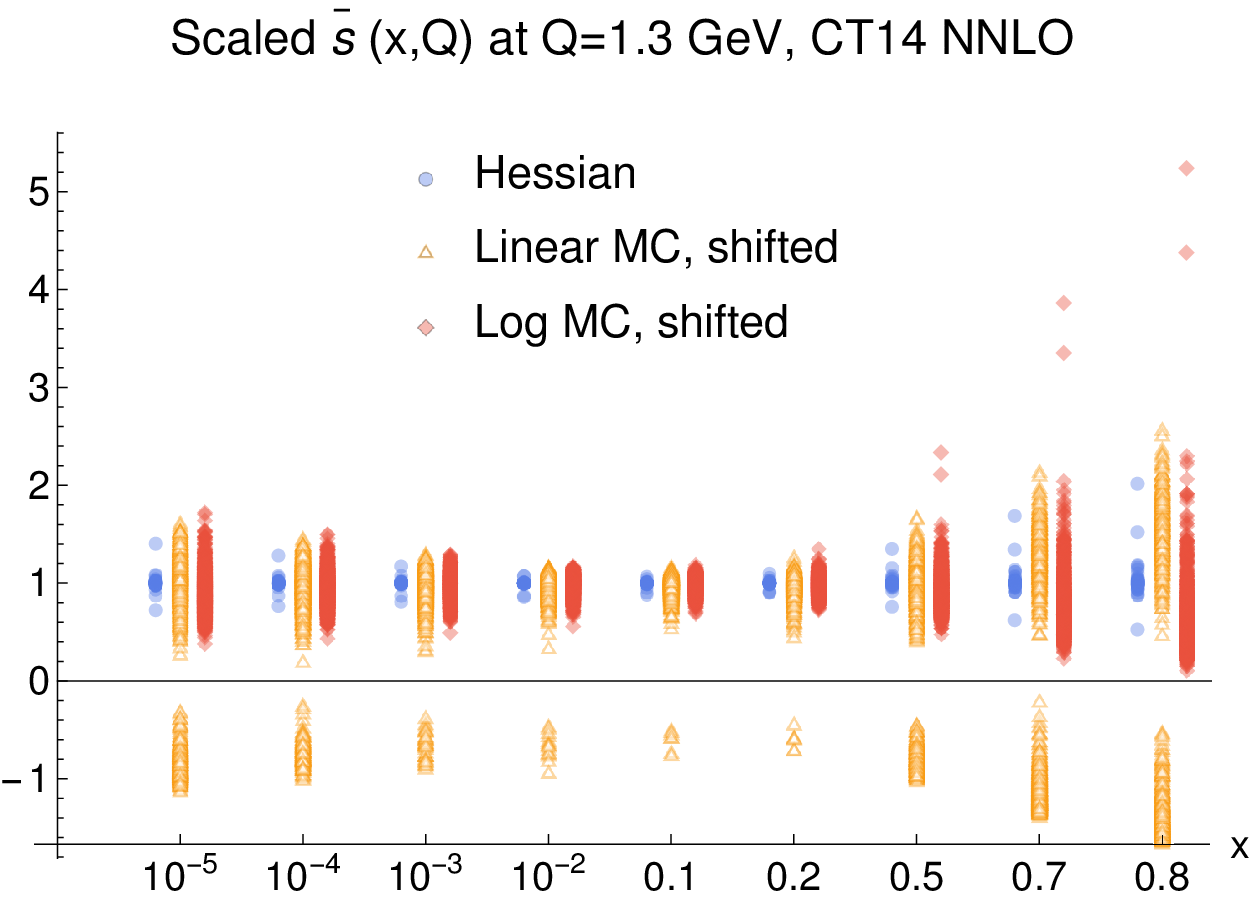}

\caption{Distributions of individual replicas for MC1 (linear MC, shifted)
and MC2 (log MC, shifted) ensembles. \label{fig:CT14MCreplicas}}
\end{figure}

The difference between the
two types of replicas is illustrated by Fig.~\ref{fig:CT14MCreplicas},
showing PDFs for the CT14 NNLO 56 Hessian eigenvector sets (blue circles)
and 1000 MC replicas obtained with linear sampling (orange triangles)
and logarithmic sampling (red diamonds). The MC replicas are shifted
as described above to have $\langle f\rangle_{MC}$ equal to $f_{0}$,
the central PDF of the Hessian set. The vertical axis is scaled as
$\left|f/f_{0}\right|^{0.2}\mbox{sign}\left(f\right)$ to better visualize
relative variations of both signs in an extended magnitude range.

The value of 1 corresponds to the replica PDF coinciding with that
of the central set. The majority of replicas produce PDF values that are close
to the central PDF set, within the intervals enclosing 1 where many 
replica symbols overlap. Some replicas for less constrained PDFs
produce very large deviations, corresponding to the values far from
1. The Hessian and log-sampled replicas are non-negative by construction,
facilitating positivity of physical cross sections. The linearly sampled
replicas can take negative values in extrapolation regions (i.e., lie below
the horizontal line at zero). In practice, it is desirable to have
non-negative cross sections for individual replicas, even though the
uncertainty band does extend to zero at some confidence level.

\clearpage{}
\section{Comparisons of replica sets and Hessian error PDFs}

\label{sec:sec3}

\subsection{Estimation of PDF uncertainties}

To justify the use of the replica sets for estimating the PDF uncertainties,
we must first show that the mean and standard deviation of the replica
PDFs themselves are consistent with the results obtained directly
from the Hessian error PDFs. A complete set of the comparisons is available
at \cite{CT14MCwebsite}. The PDFs are generated by sampling LHAPDF6
grid files using the {\sc mcgen} code \cite{MCGENwebsite}. 

\begin{figure}
\includegraphics[width=0.49\textwidth]{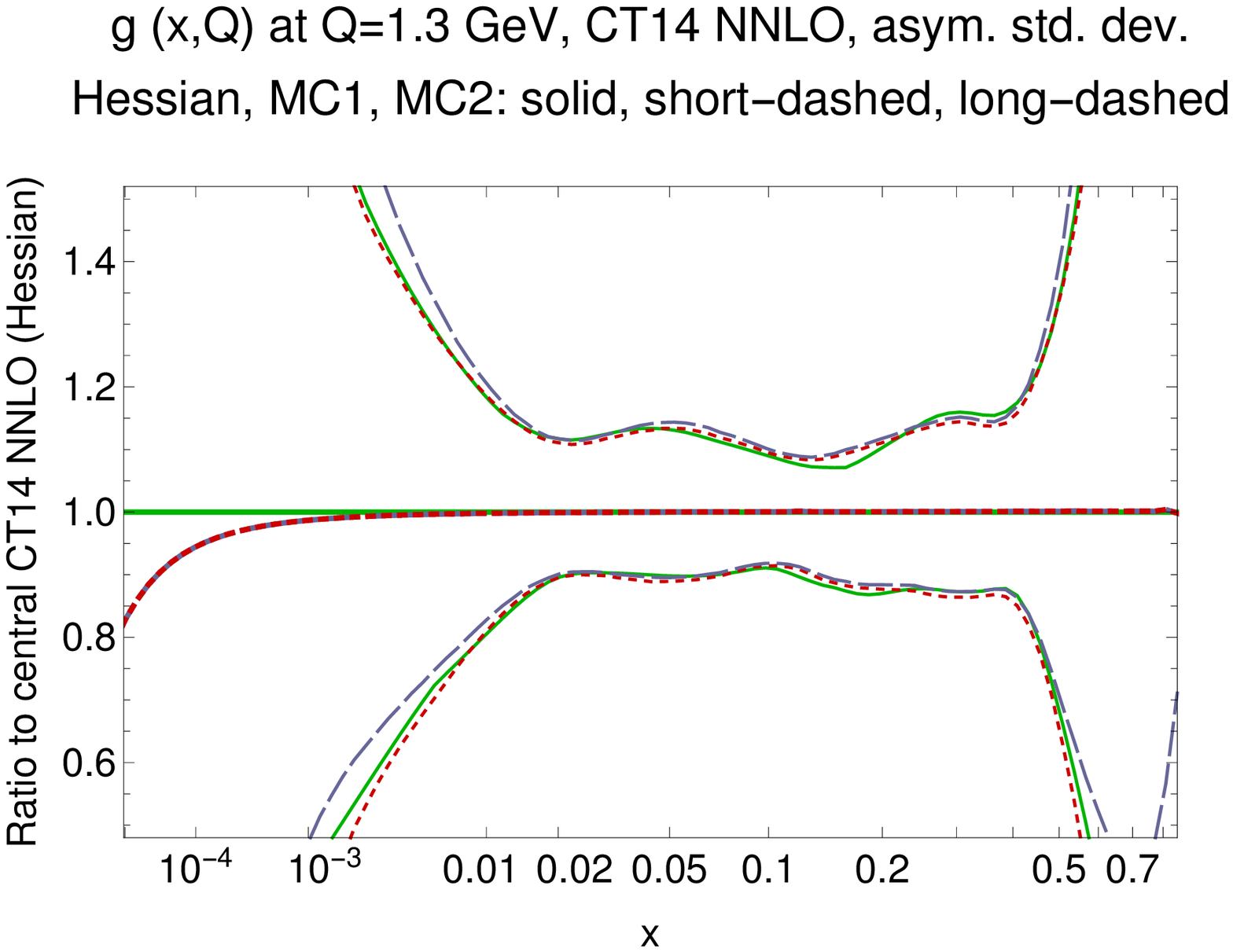}\includegraphics[width=0.49\textwidth]{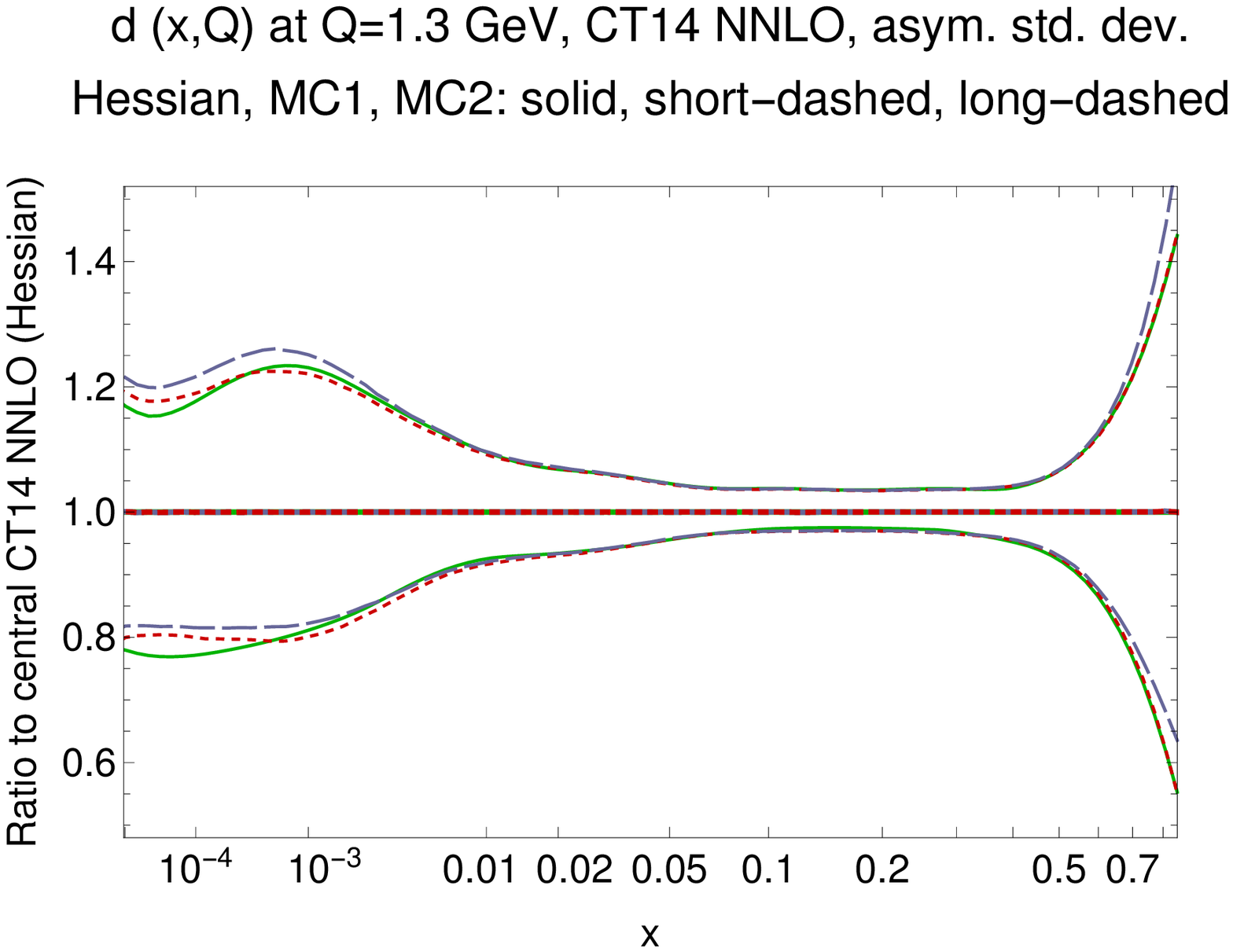}

\includegraphics[width=0.49\textwidth]{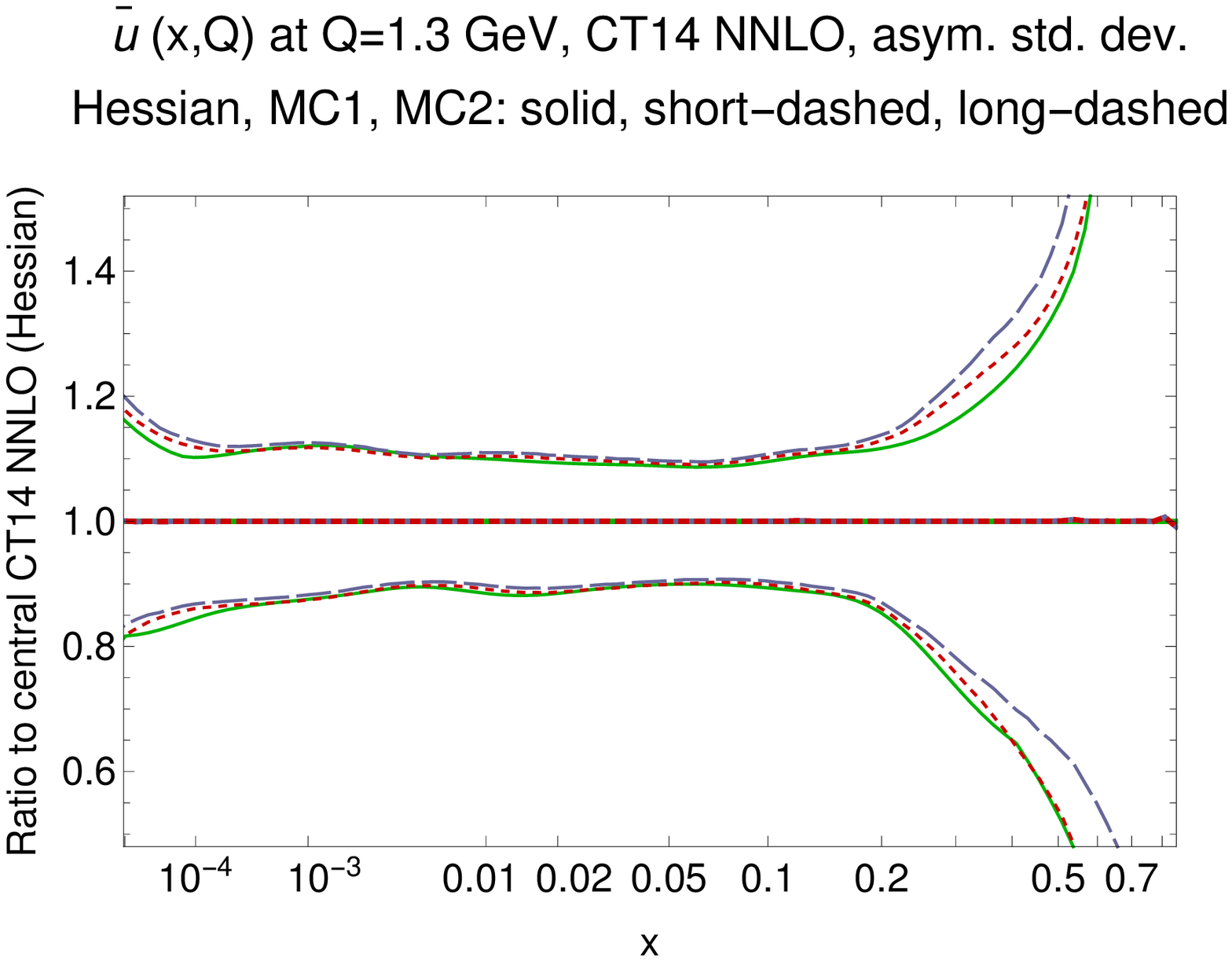}\includegraphics[width=0.49\textwidth]{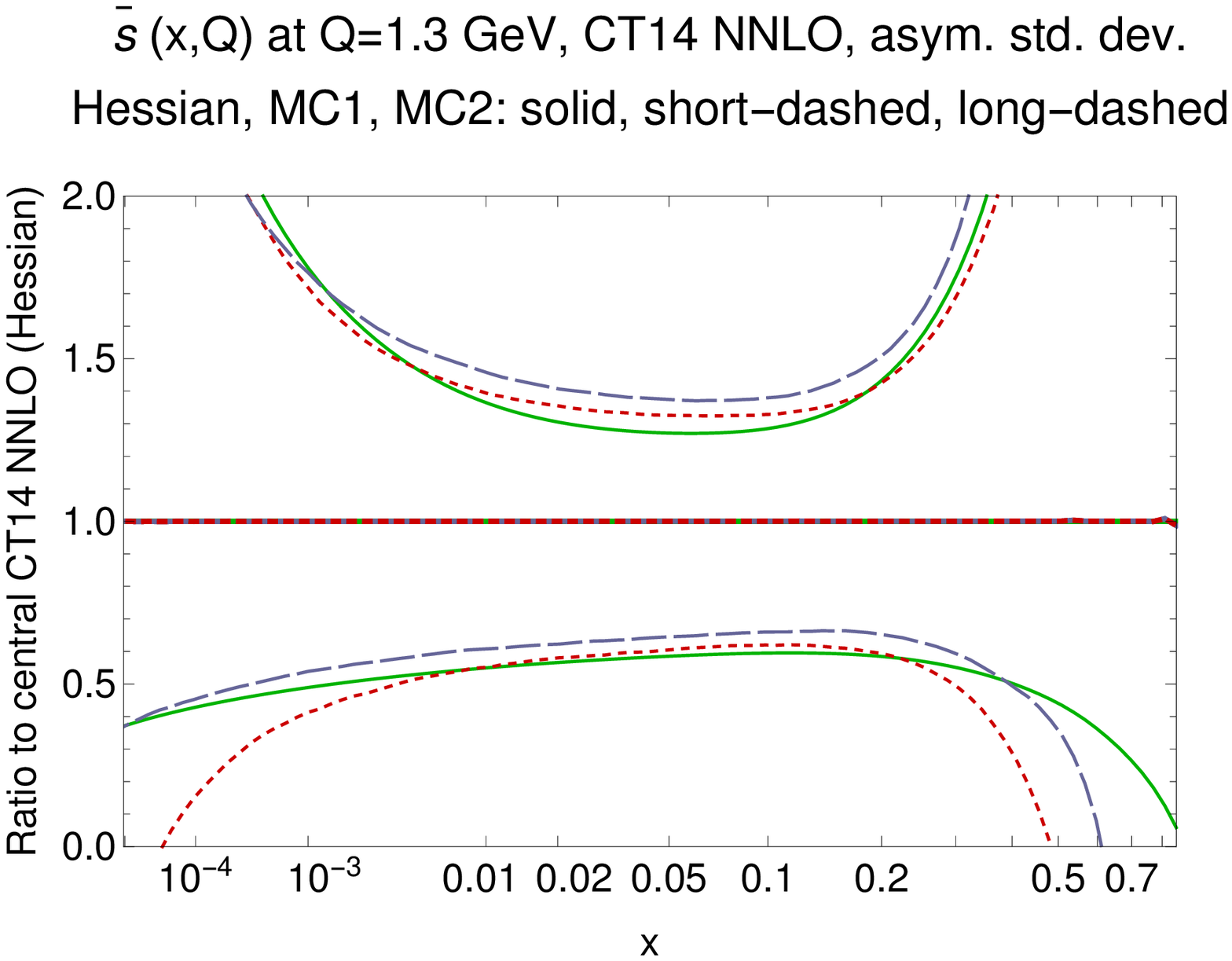}

\caption{The mean values and asymmetric standard deviations (\ref{dMCXp}),
(\ref{dMCXm}) of the CT14 NNLO MC1 (short-dashed) and MC2 (long-dashed)
PDFs, compared to the mean and 68\% c.l. uncertainty (Eq.~(\ref{dH68X}),
solid) of the CT14 NNLO Hessian PDF. The PDFs are shown as ratios
to the central CT14 fit. Upper panel: $g(x,Q_{0})$ and $d(x,Q_{0})$.
Lower panel: $\bar{u}(x,Q_{0})$ and $\bar{s}(x,Q_{0})$. \label{fig:CT14MCasymNNLO}}
\end{figure}

Figure \ref{fig:CT14MCasymNNLO} shows the values and error bands, for
the ratios of $g(x,Q_{0})$, $d(x,Q_{0})$, $\bar{u}(x,Q_{0}),$ and
$\bar{s}(x,Q_{0})$ PDFs to the respective central CT14 PDFs, at the
initial scale $Q_{0}=1.3$ GeV, at NNLO. The three green solid curves
represent the central PDF (for which the ratio is 1), and the asymmetric
upward and downward 68\% c.l. deviations calculated from the Hessian
master formulae (\ref{dH68Xp}) and (\ref{dH68Xm}). The red short-dashed
and blue long-dashed curves show the mean and the error bands corresponding
to asymmetric standard deviations of the CT14 MC1 and MC2 ensembles
with 1000 replicas. We see in the figure that the Hessian and replica
PDFs are very consistent in terms of both the means and standard values.%
\footnote{The central gluon PDF from the replica set is slightly lower than
the Hessian one in the upper left Fig.~\ref{fig:CT14MCasymNNLO}
at $x<10^{-3}$ and $Q_{0}=1.3$ GeV, where the CT14 gluon vanishes,
as a result of numerical roundoff errors. %
} Both MC1 and MC2 ensembles, obtained by linear and logarithmic sampling
of the replicas, result in agreement of comparable quality with the
Hessian PDFs at intermediate $x.$ In the extreme $x$ regions, the
MC2 error bands tend to be slightly shifted toward positive values. The differences may be somewhat more noticeable for the poorly constrained strangeness PDF, 
but in all cases the Hessian and MC error bands are fully
compatible within the uncertainties of the definition of PDF errors. 

\begin{figure}
\includegraphics[width=0.49\textwidth]{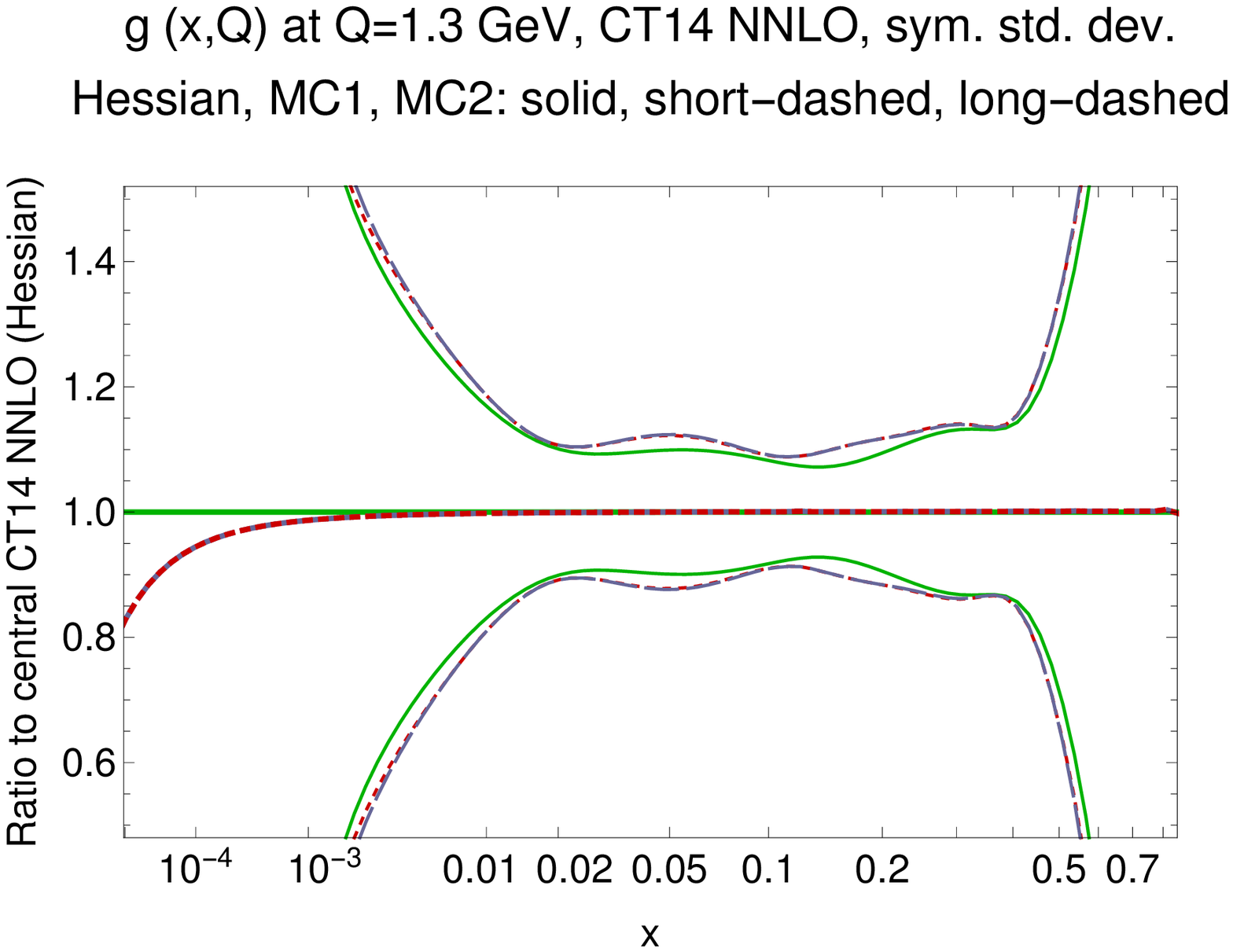}\includegraphics[width=0.49\textwidth]{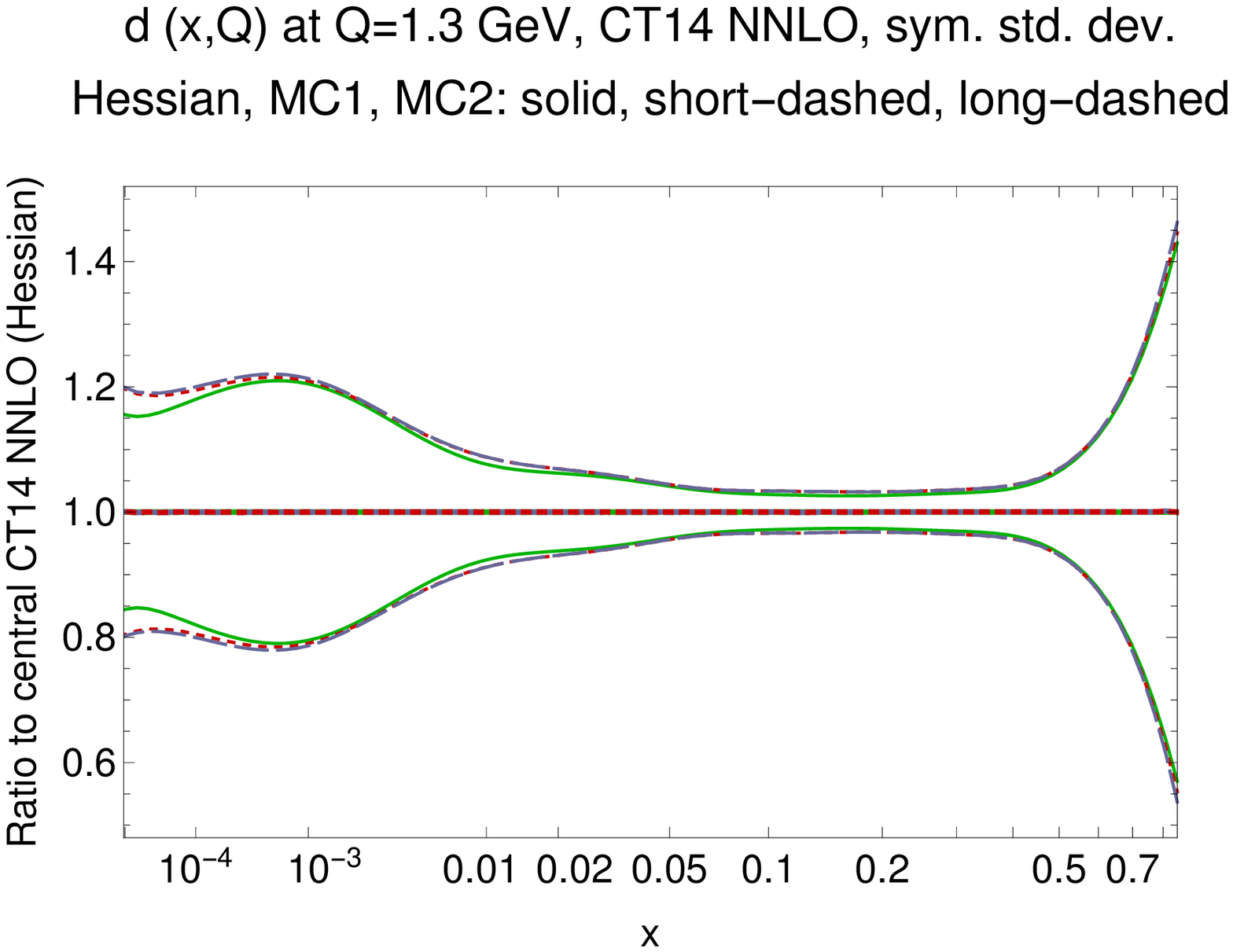}

\includegraphics[width=0.49\textwidth]{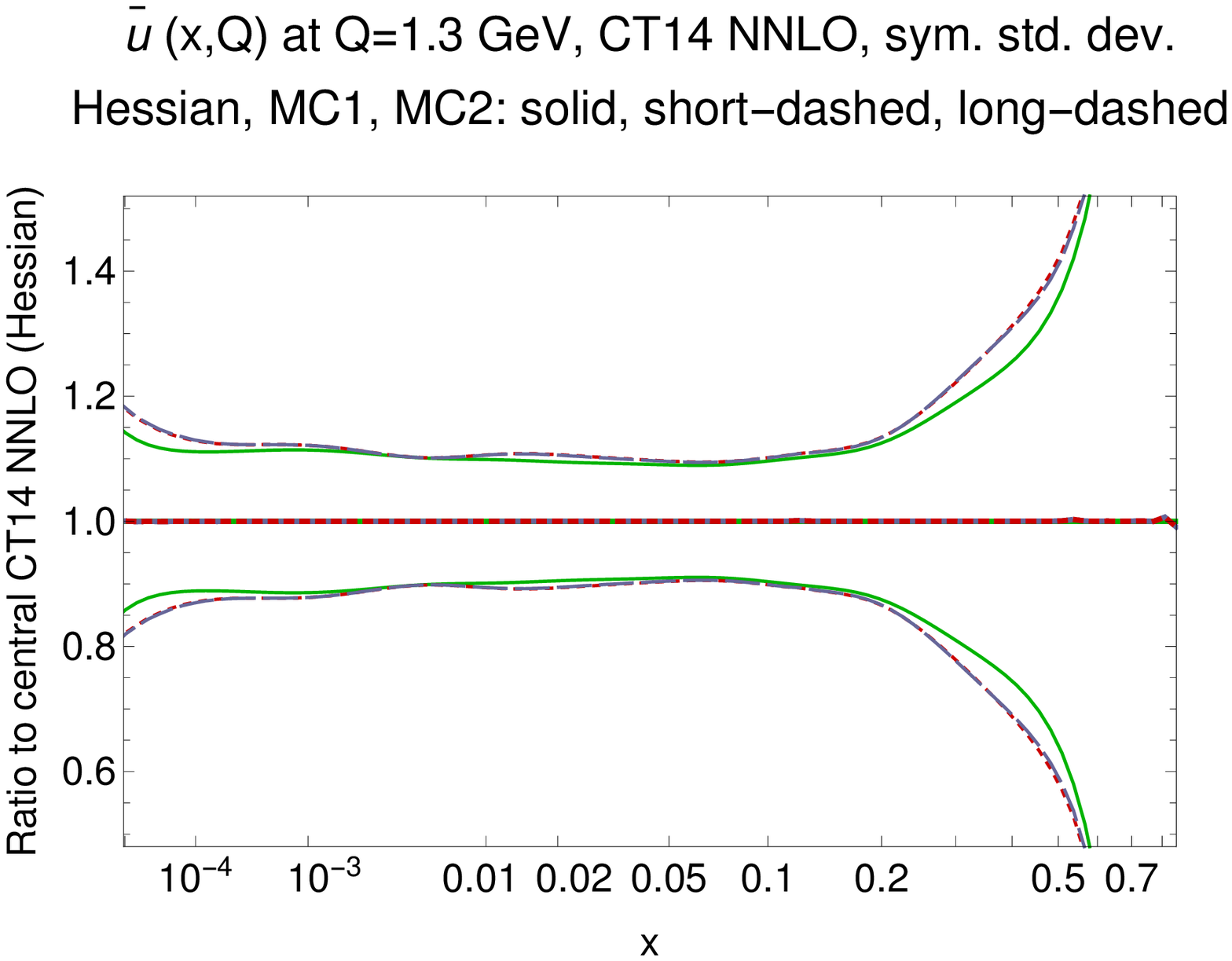}\includegraphics[width=0.49\textwidth]{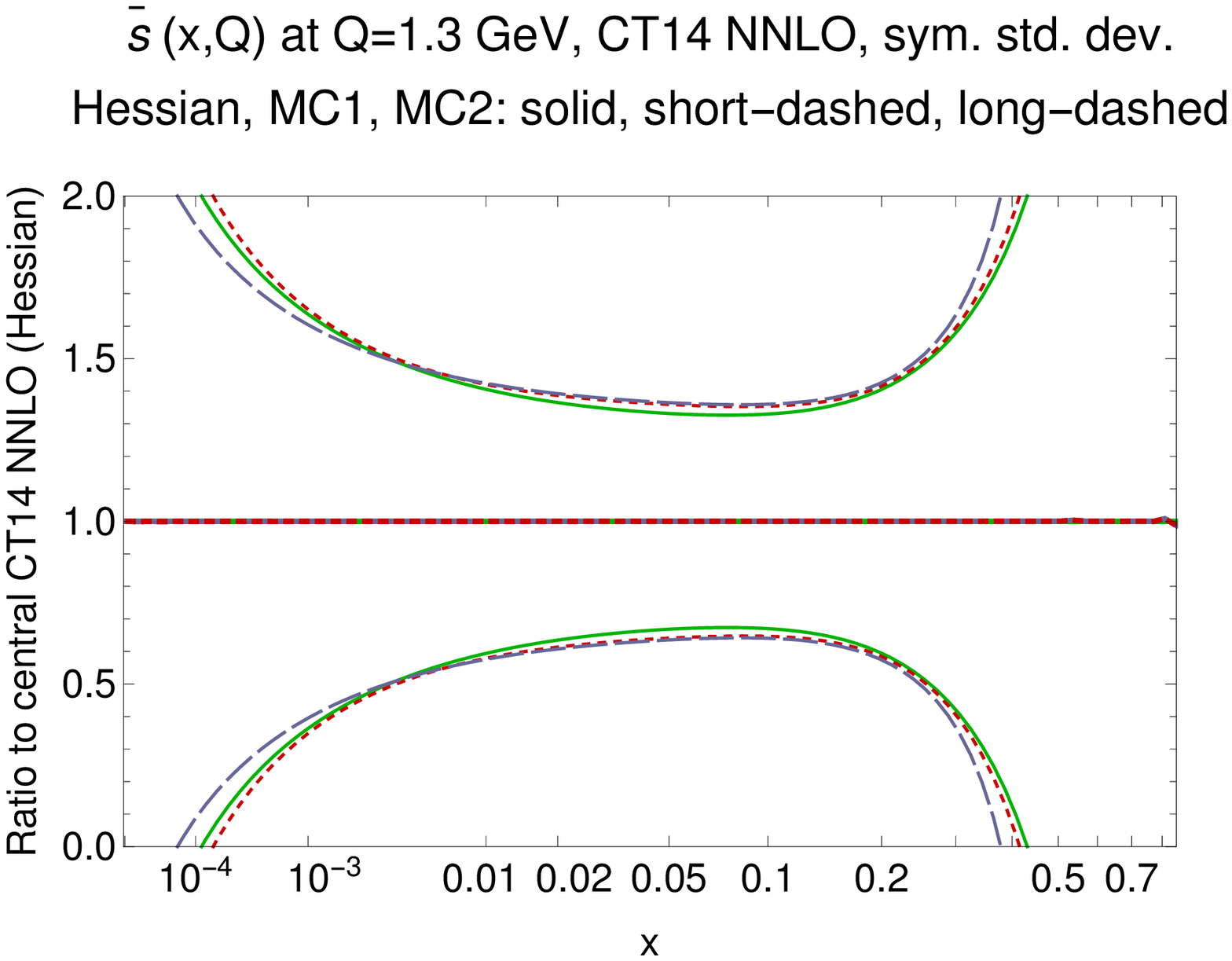}

\caption{Same as Fig.~\ref{fig:CT14MCasymNNLO}, for symmetric Hessian uncertainties
(\ref{dH68X}) and symmetric MC standard deviations (\ref{dMCX}).\label{fig:CT14MCsymNNLO}}
\end{figure}

\begin{figure}
\includegraphics[width=0.49\textwidth]{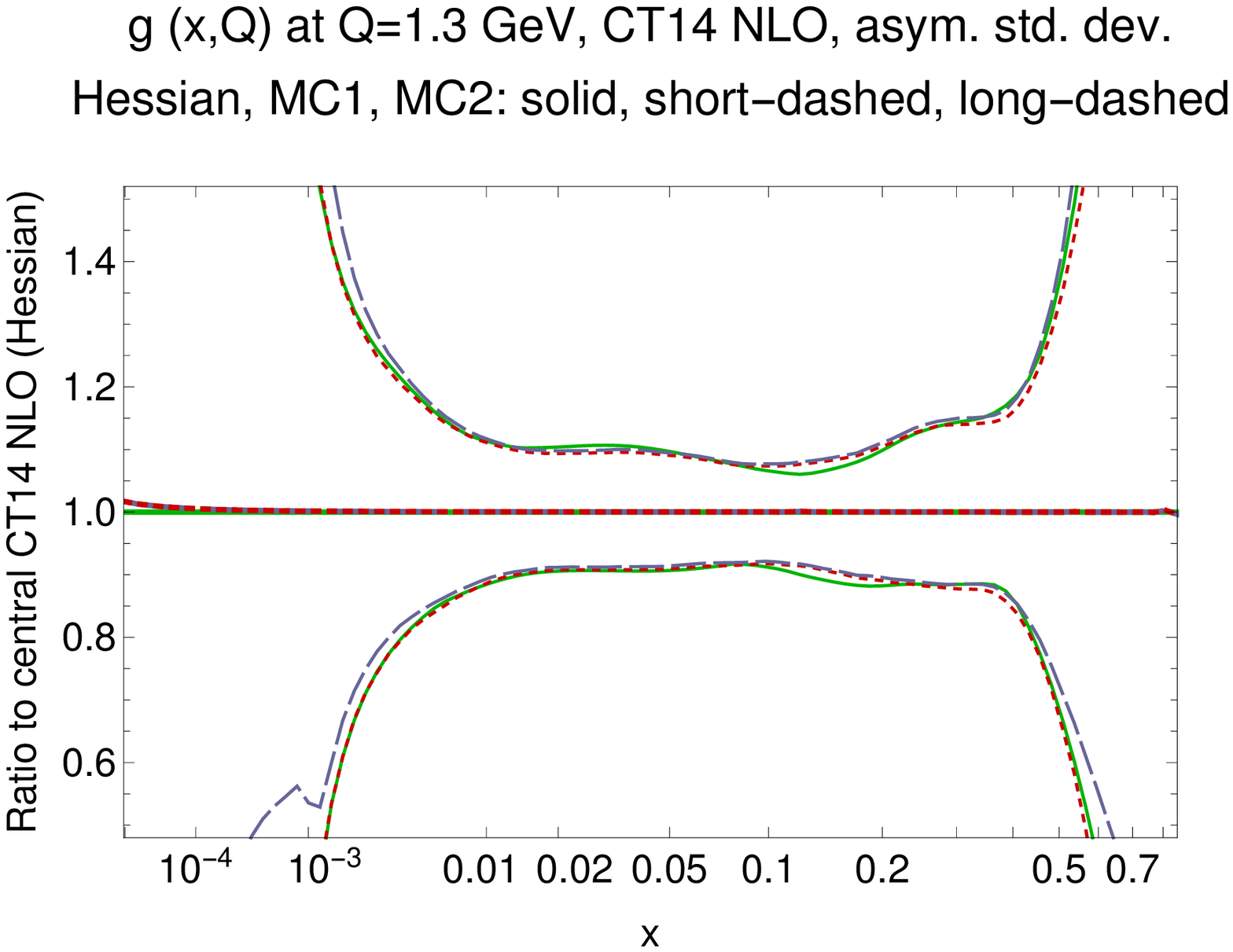}\includegraphics[width=0.49\textwidth]{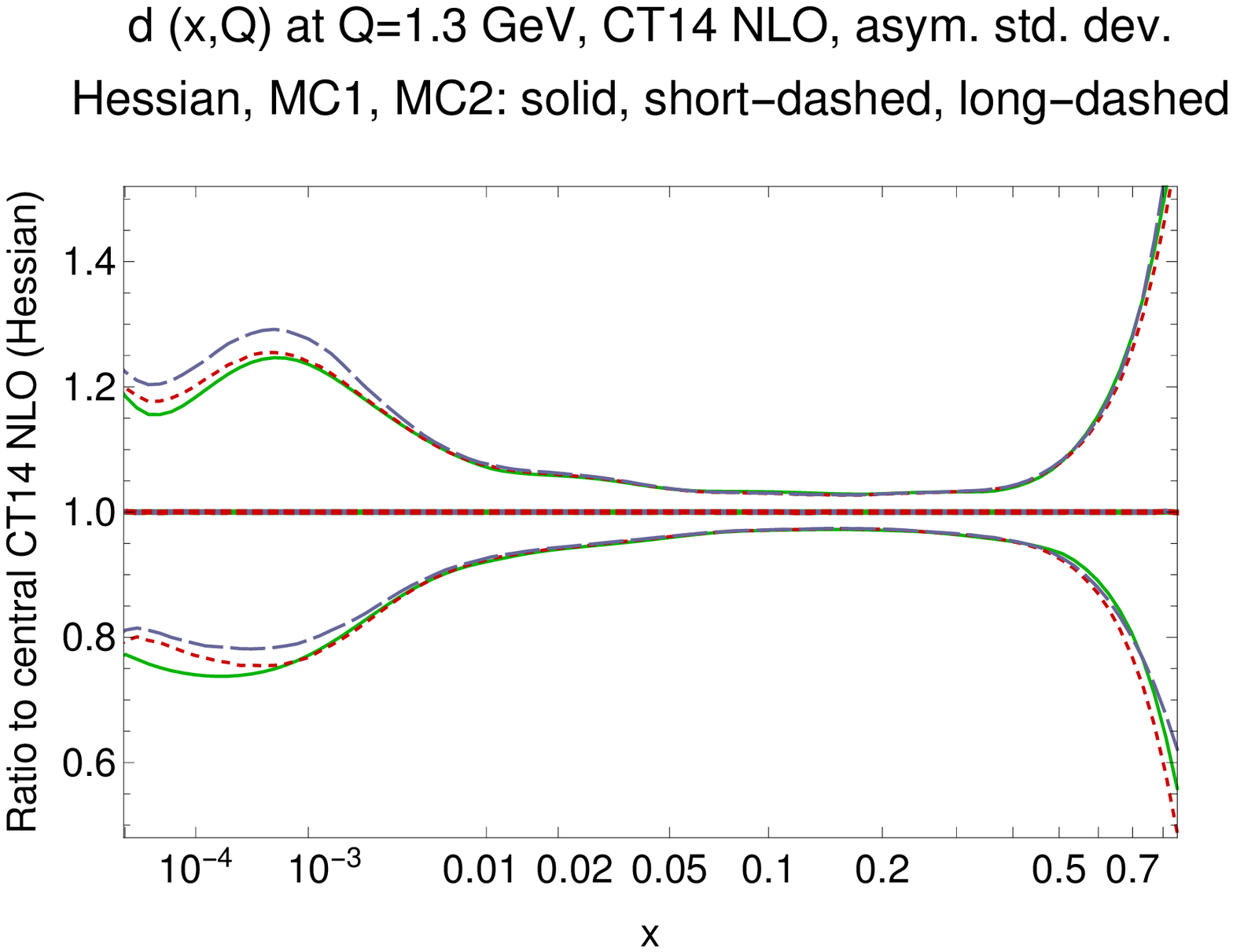}

\includegraphics[width=0.49\textwidth]{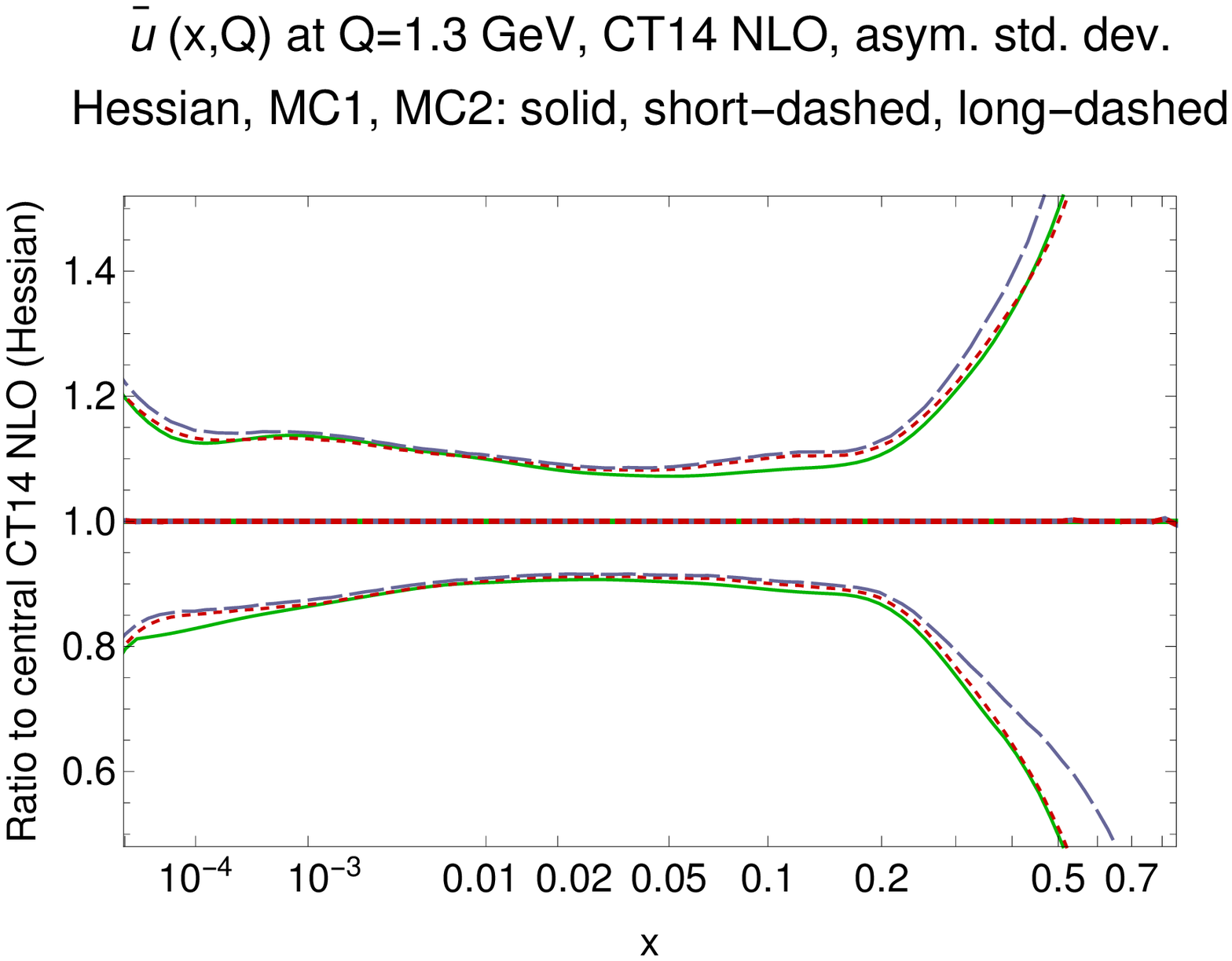}\includegraphics[width=0.49\textwidth]{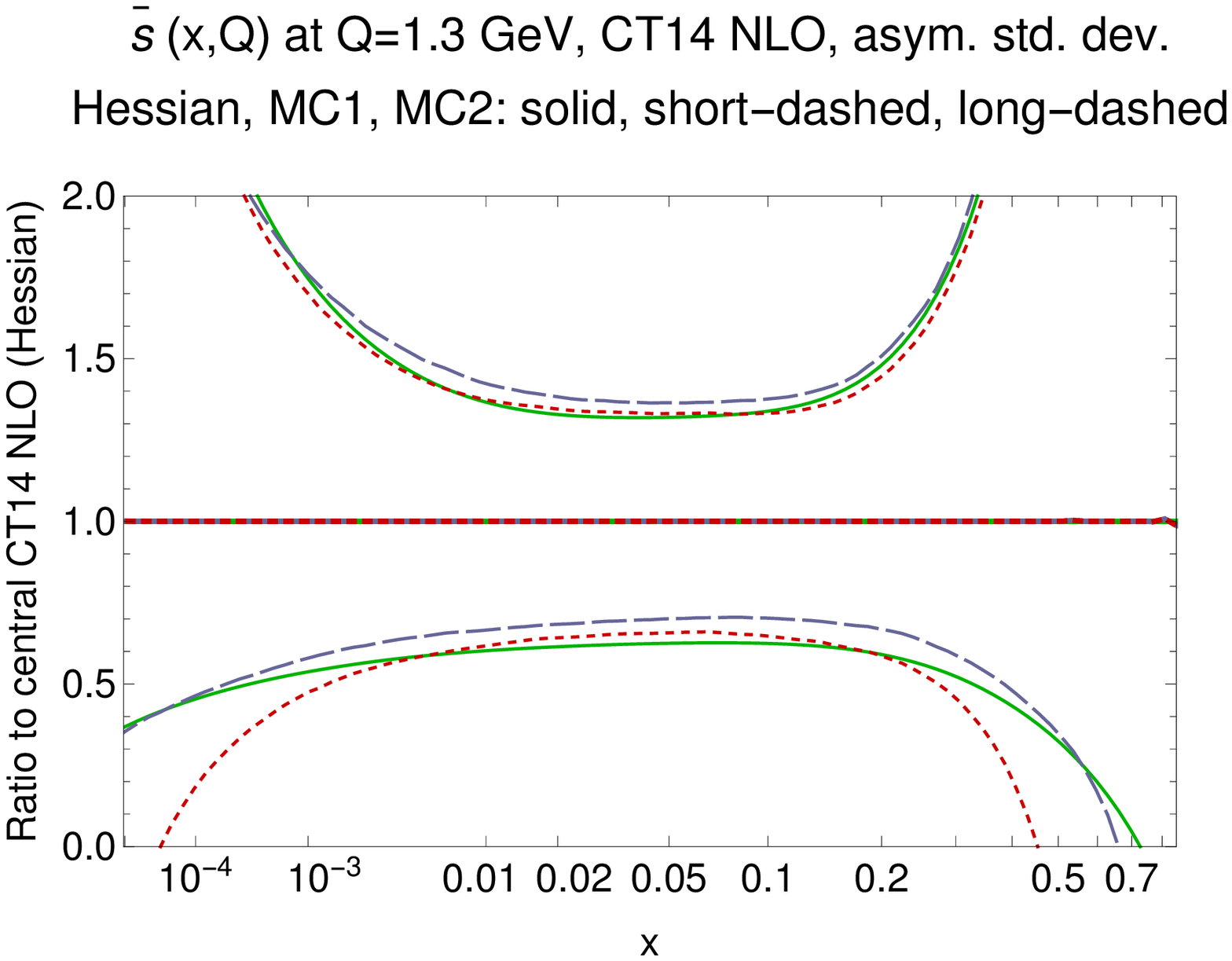}

\caption{Same as Fig.~\ref{fig:CT14MCasymNNLO}, at NLO. \label{fig:CT14MCasymNLO}}
\end{figure}

In the case of symmetric Hessian uncertainties (\ref{dH68X}) and
symmetric standard deviations (\ref{dMCX}), the agreement of Hessian,
MC1, and MC2 NNLO ensembles is even better. The corresponding comparisons
are shown in Fig.~\ref{fig:CT14MCsymNNLO}. Equivalent comparisons
for CT14 NLO asymmetric errors are presented in Fig.~\ref{fig:CT14MCasymNLO}.
These figures compare the parton distribution functions at the initial
momentum scale $Q_{0}$. If we evolve these to any $Q>Q_{0}$ using
the DGLAP equations, the agreement between the three ensembles further
improves. Figures with comparisons of the CT14 Hessian, MC1, and MC2 PDFs
at $Q=2$ and $100$ GeV can be viewed at \cite{CT14MCwebsite}.

\begin{figure}
\includegraphics[width=0.49\textwidth]{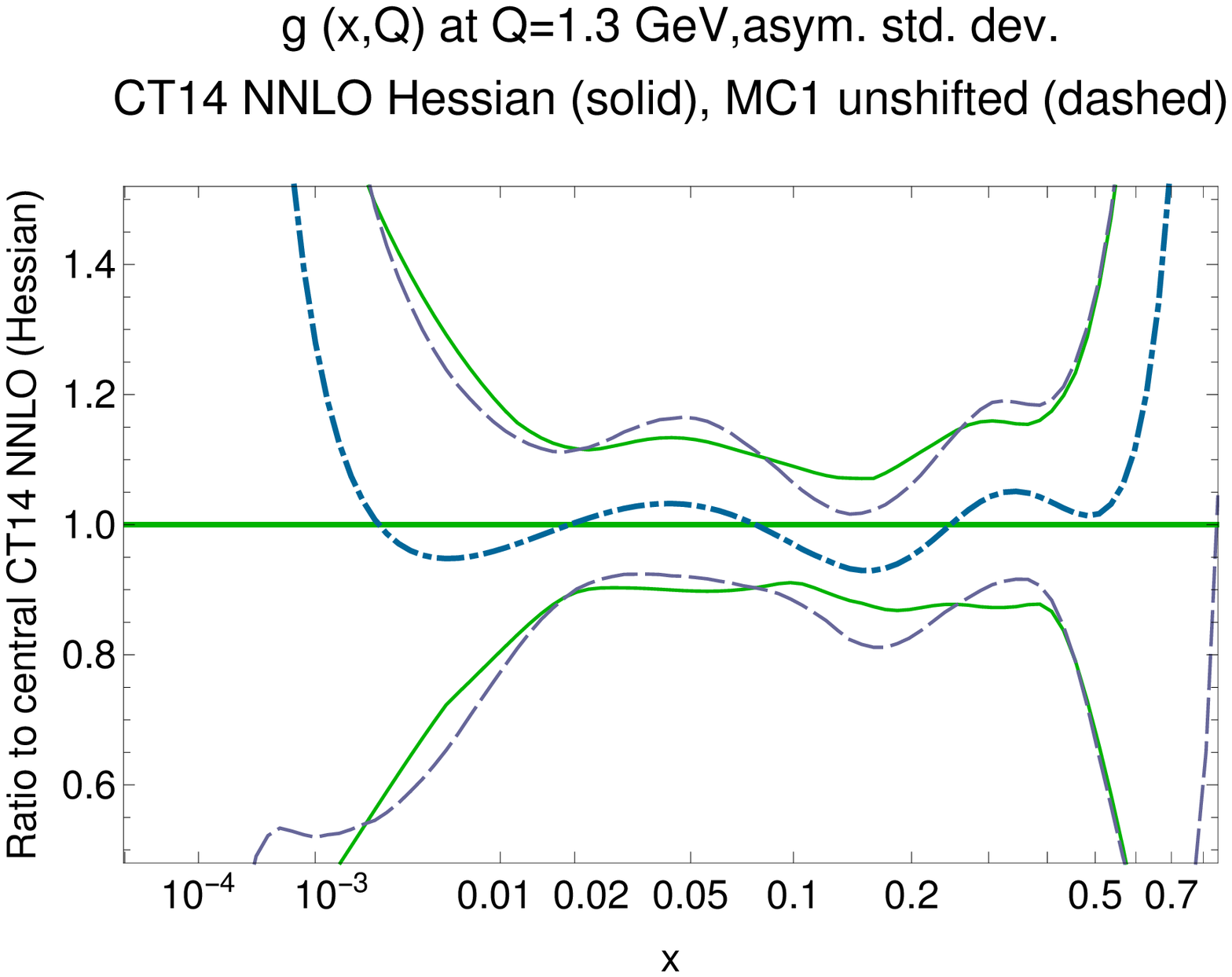}\includegraphics[width=0.49\textwidth]{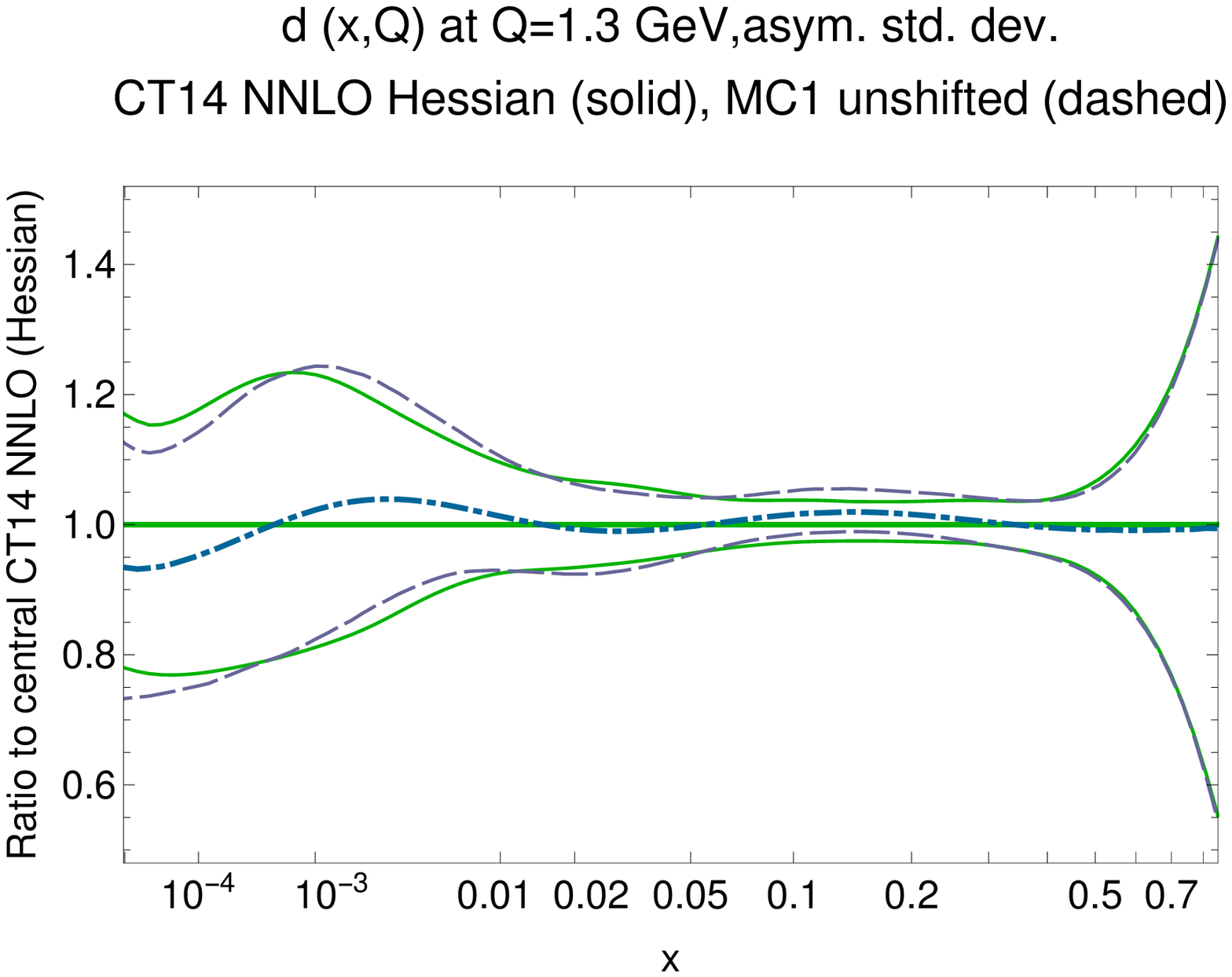}

\includegraphics[width=0.49\textwidth]{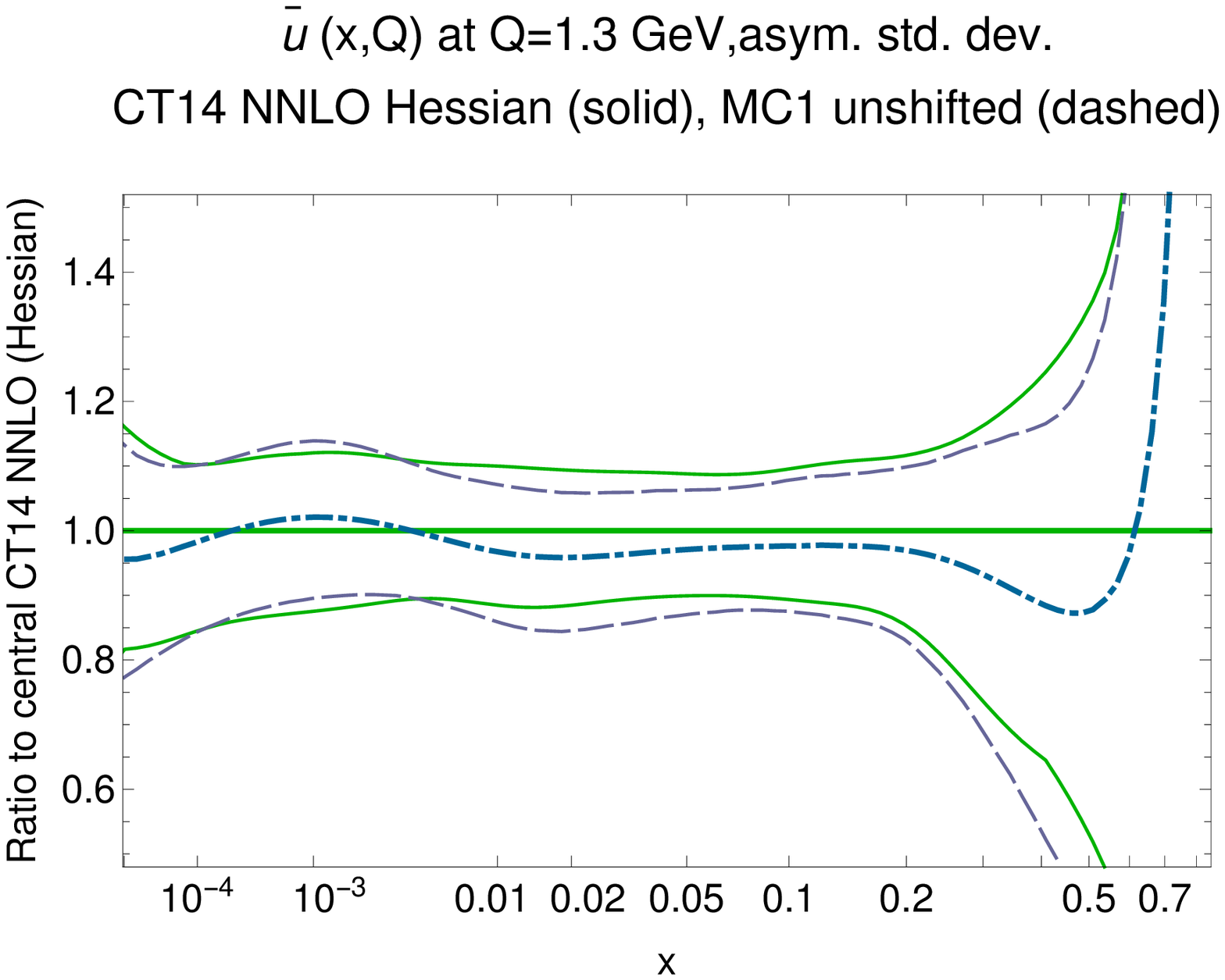}\includegraphics[width=0.49\textwidth]{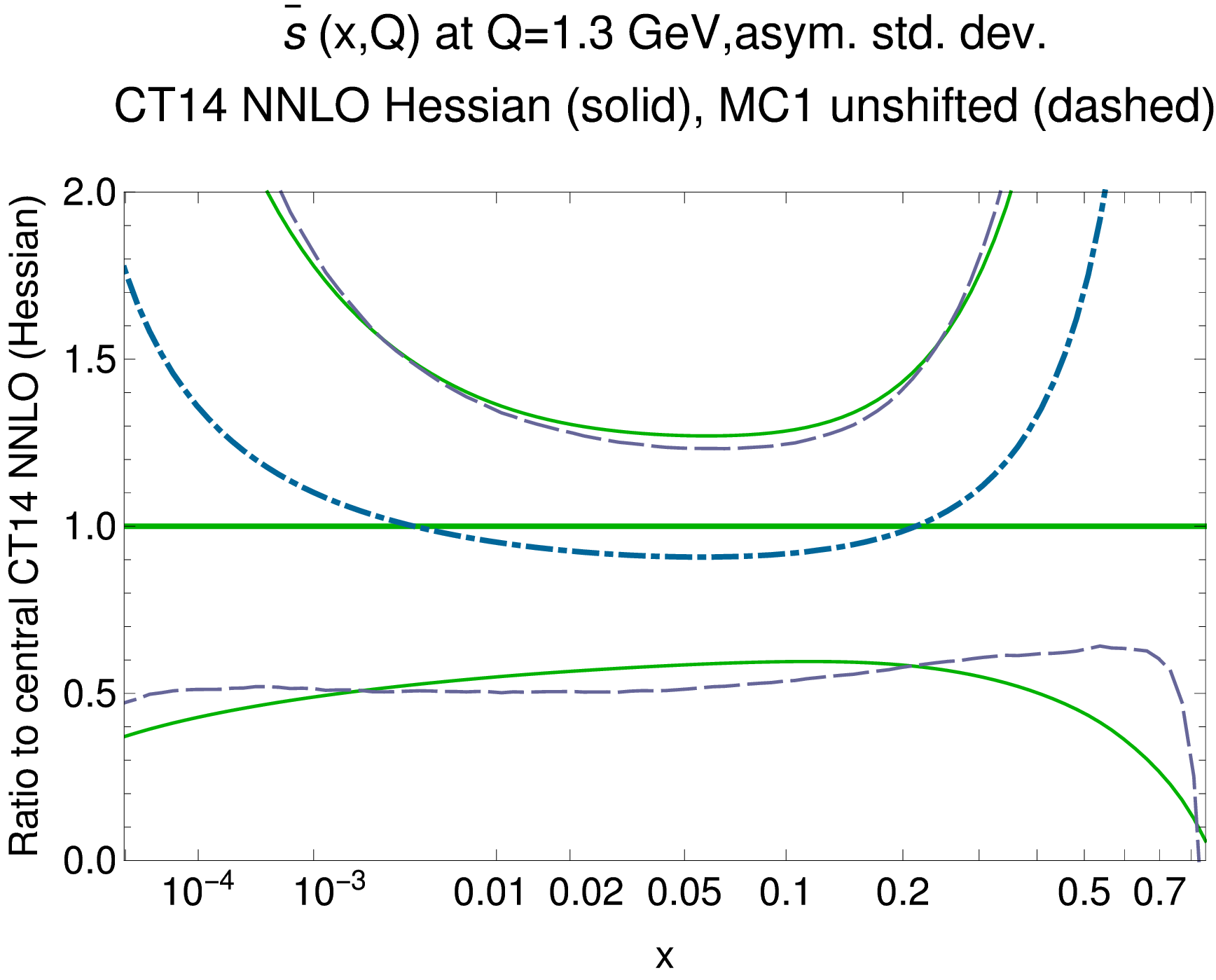}

\caption{Same as Fig.~\ref{fig:CT14MCasymNNLO}, for unshifted MC1 and Hessian
PDFs of the CT14 NNLO family. \label{fig:CT14MCasymNNLOunshifted}}
\end{figure}

Lastly, in Fig.~\ref{fig:CT14MCasymNNLOunshifted}, we compare the
CT14 NNLO Hessian PDFs to a MC1 version that does not shift the
replicas by a constant amount $\Delta$. In this case, the whole bands
are shifted compared to the Hessian band across all $x$, 
by following their wiggly mean PDFs. [The reader
  can generate the unshifted replica ensemble with the {\sc mcgen}
  program if desired.] As was argued in the previous section, the
$\Delta$ shift is introduced to eliminate such spurious variations
introduced by truncation of the Taylor series.

It is also useful to compare the {\em luminosity functions}
and their uncertainties, for the MC replicas and the Hessian PDFs.
The luminosity function of a parton-parton pair, for production of
a final state $M^{2}$ at collider energy $\sqrt{s}$, is defined
as in \cite{Campbell:2006wx}, with an additional constraint that
the rapidity of the final state, given by $y = \frac{1}{2} \ln (x_{2}/x_{1})$,
does not exceed $y_{cut}$ in magnitude:
\begin{equation}
L_{ab}(s,M^{2},y_{cut})=\frac{1}{1+\delta_{ab}}\left[\int_{\frac{M}{\sqrt{s}}e^{-y_{cut}}}^{\frac{M}{\sqrt{s}}e^{y_{cut}}}\frac{d\xi}{\xi}f_{a}(\xi,M)f_{b}\left(\frac{M}{\xi\sqrt{s}},M\right)+\left(a\leftrightarrow b\right)\right].
\end{equation}

\begin{figure}
\includegraphics[width=0.49\textwidth]{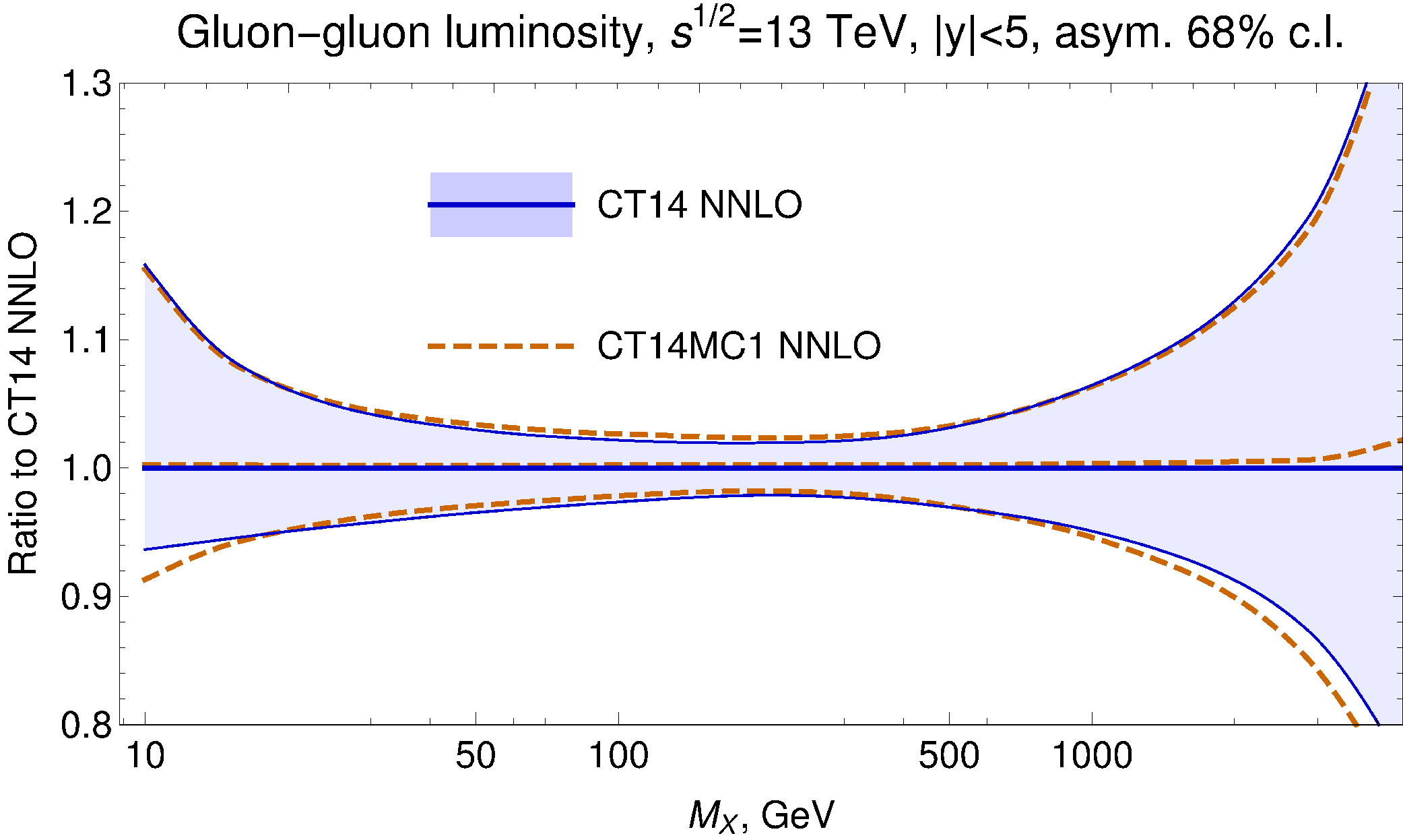}\includegraphics[width=0.49\textwidth]{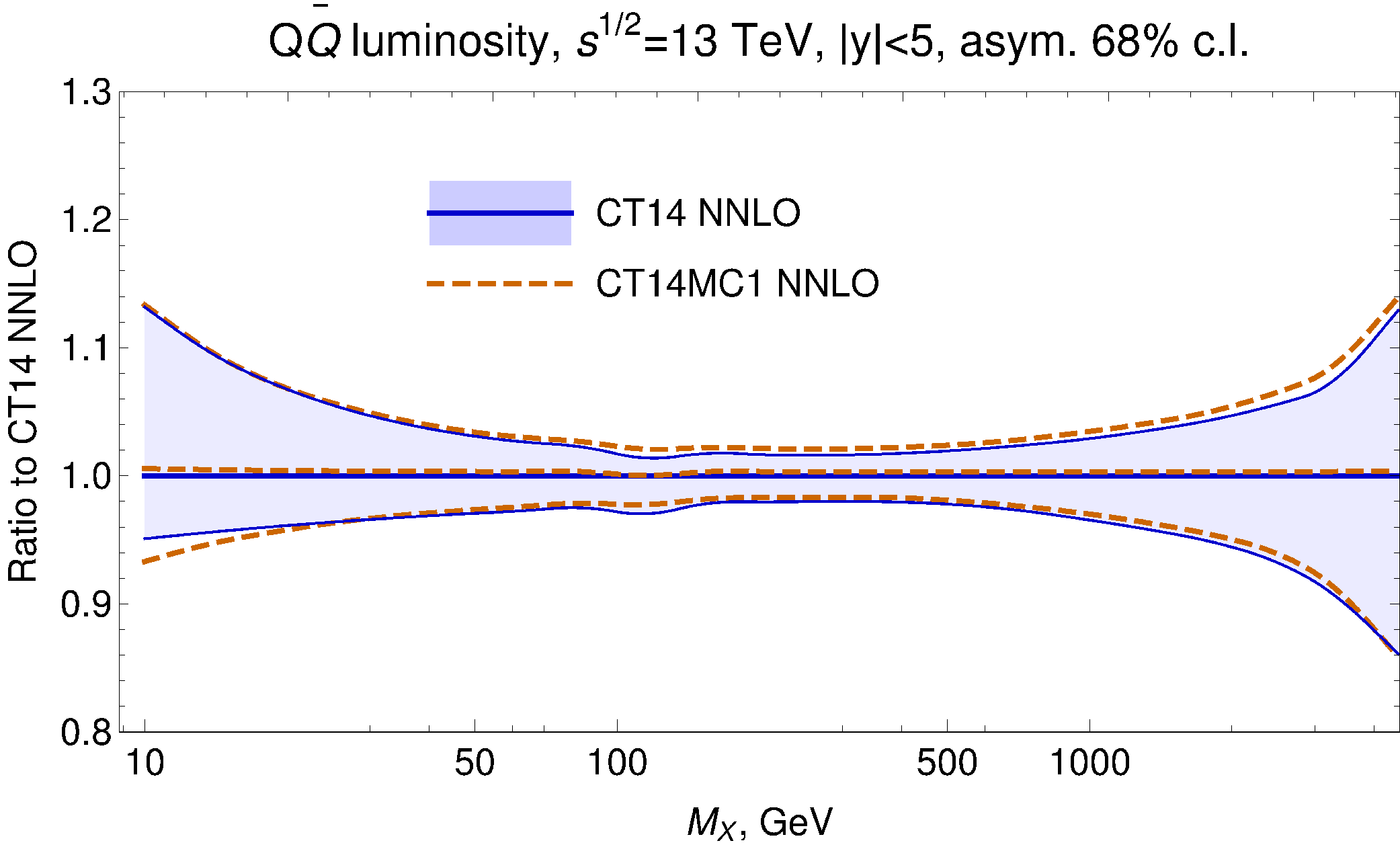}

\includegraphics[width=0.49\textwidth]{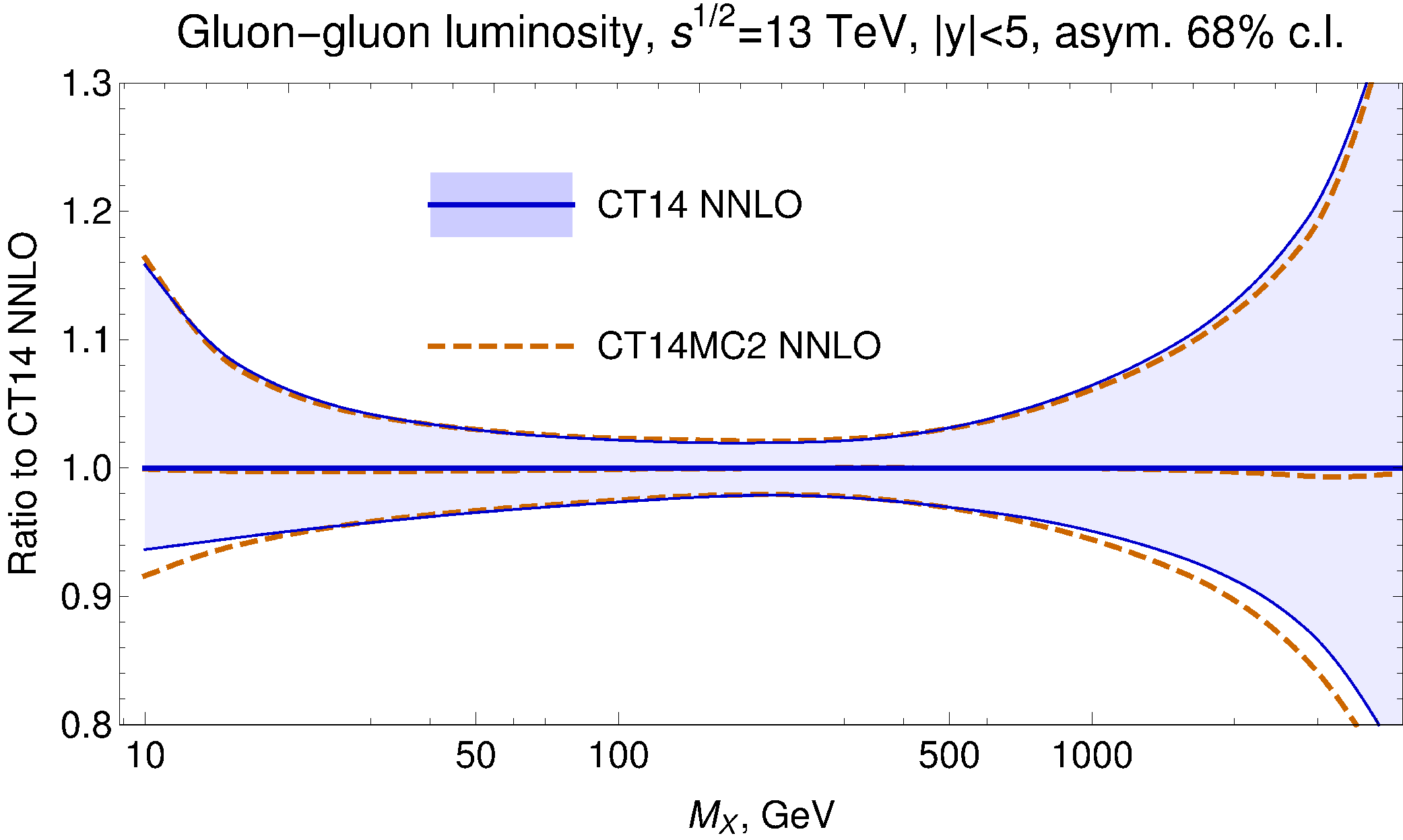}\includegraphics[width=0.49\textwidth]{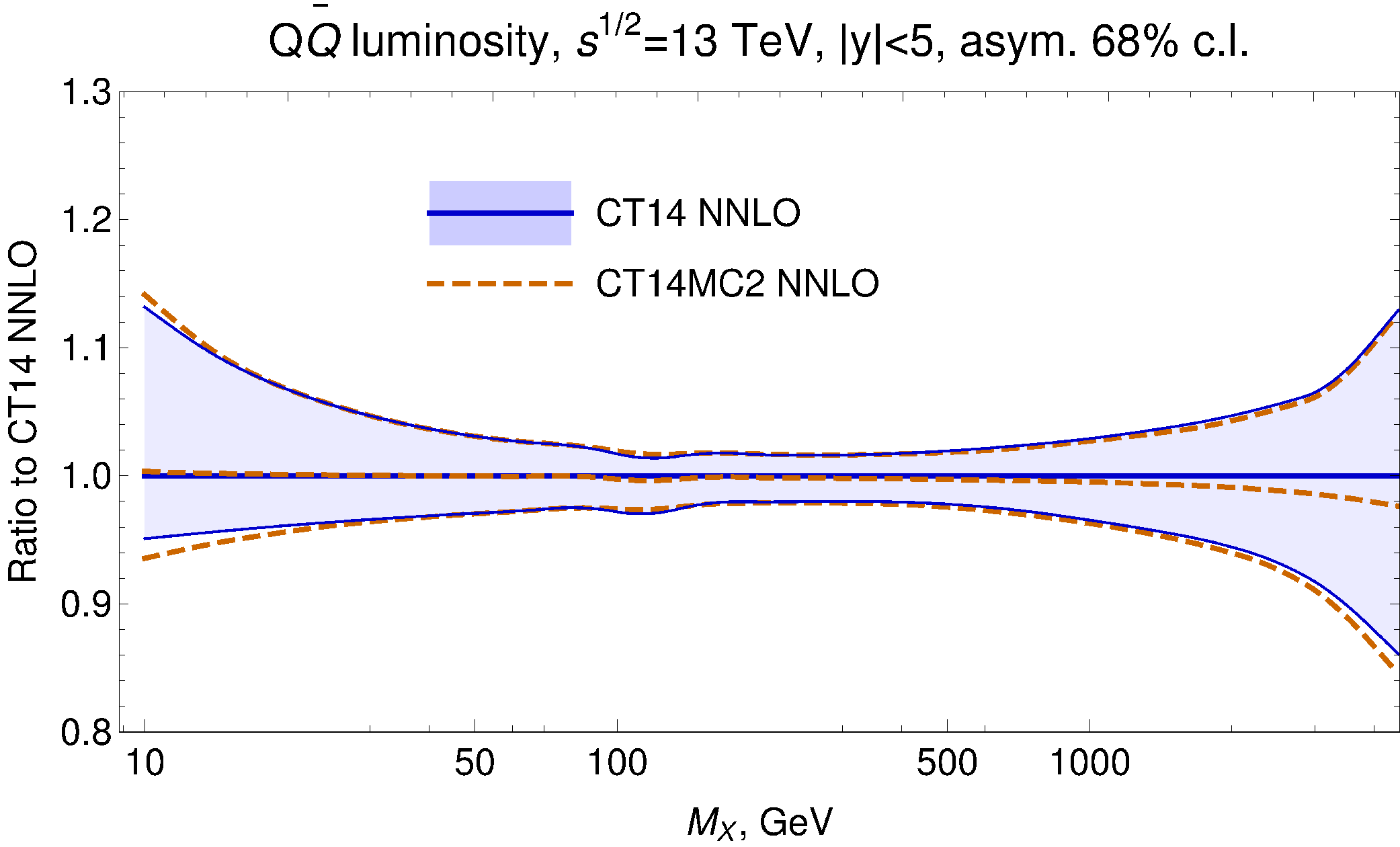}

\caption{$gg$ and $q\bar{q}$ parton luminosities for the CT14 Hessian (solid
blue) and MC (brown dashed) PDF families at NNLO, shown as the ratio
to the central Hessian PDF. \label{fig:CT14MClumiNNLO}}
\end{figure}

\begin{figure}
\includegraphics[width=0.49\textwidth]{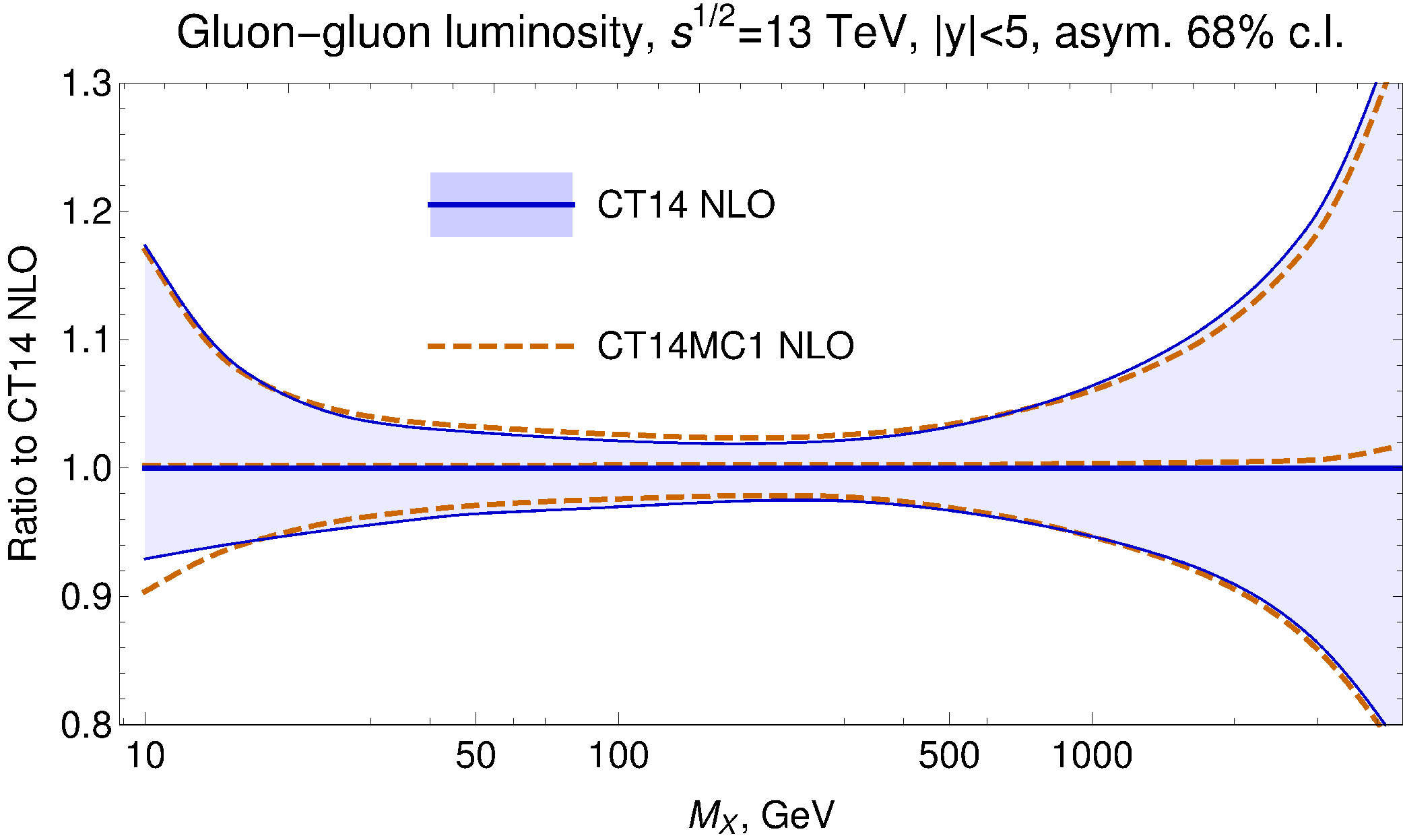}\includegraphics[width=0.49\textwidth]{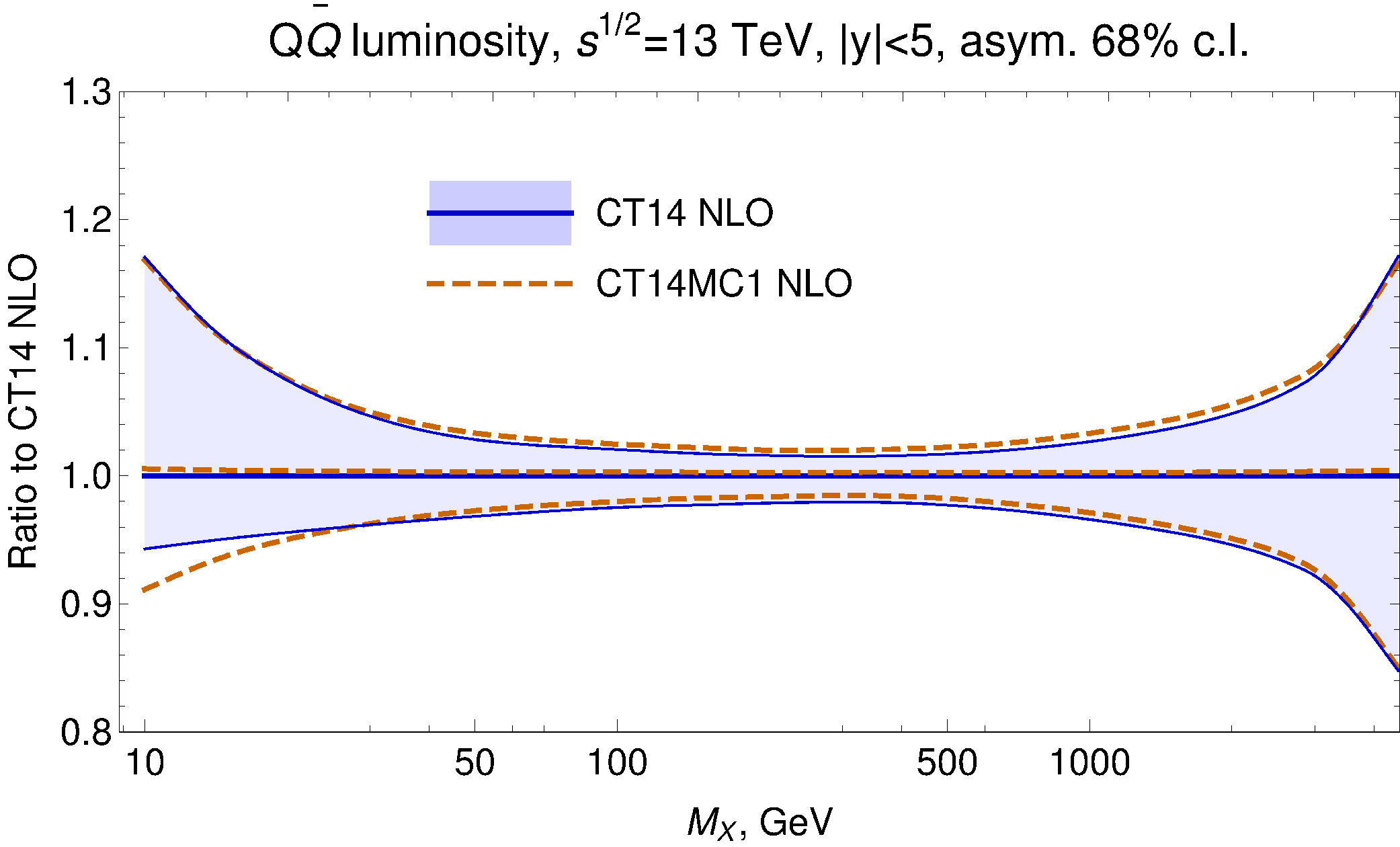}

\includegraphics[width=0.49\textwidth]{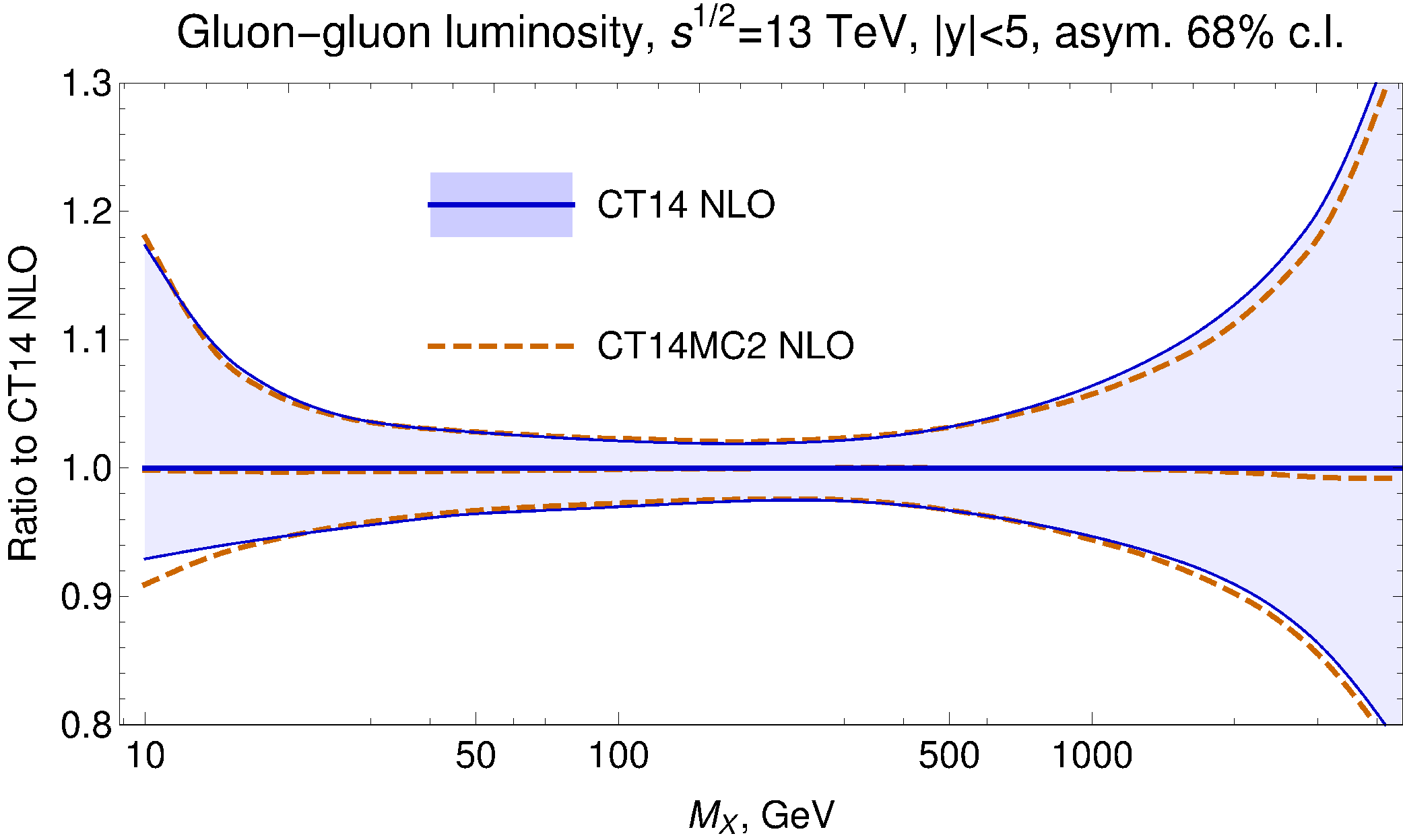}\includegraphics[width=0.49\textwidth]{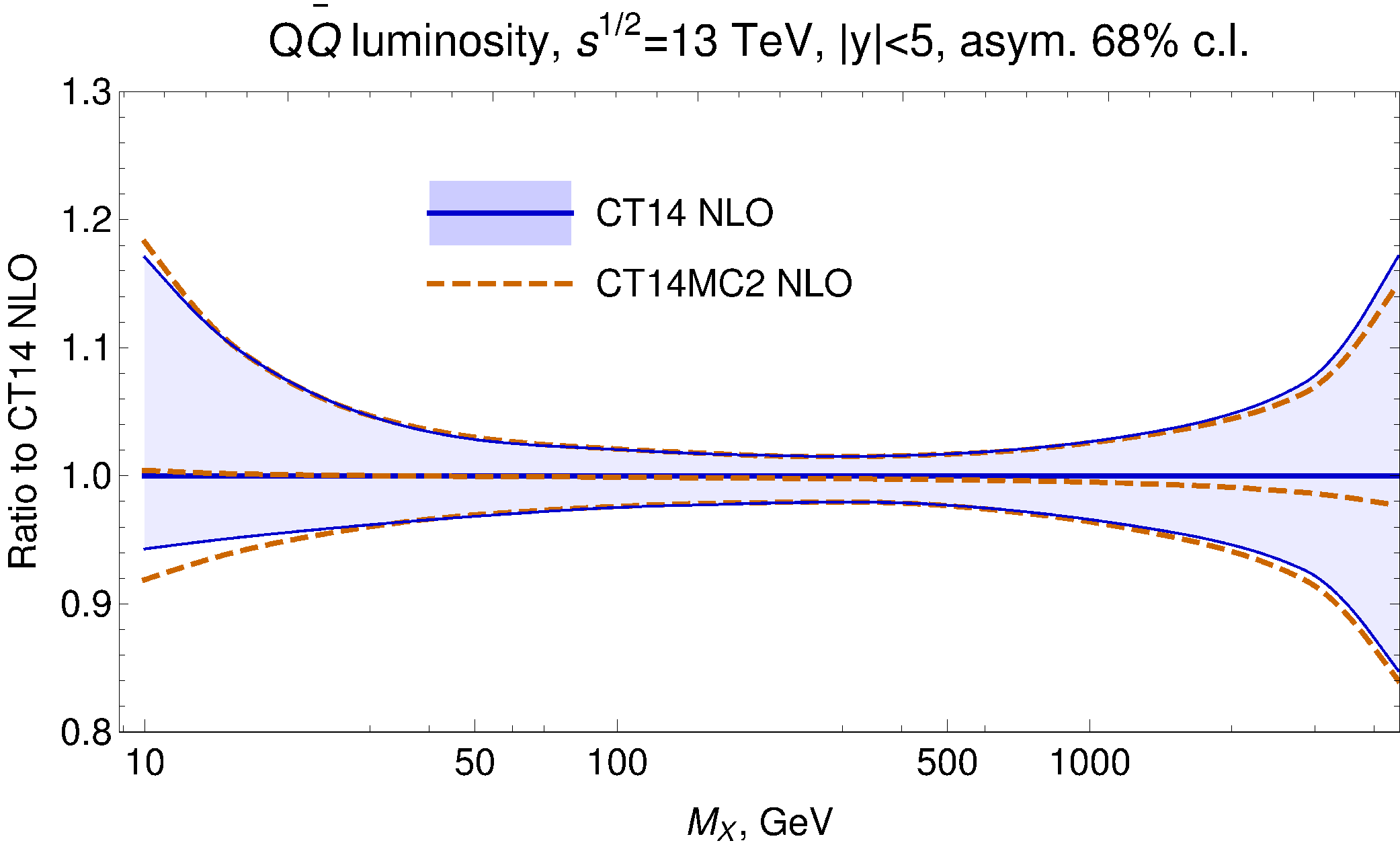}

\caption{Same as Fig.\ref{fig:CT14MCasymNLO}, for NLO luminosities. \label{fig:CT14MClumiNLO}}
\end{figure}

The gluon-gluon and quark-antiquark luminosity functions, calculated
from the CT14 Hessian and MC PDFs according to this formula, are shown
in Fig. \ref{fig:CT14MClumiNNLO} and \ref{fig:CT14MClumiNLO} as
functions of $M^{2}$ with $\sqrt{s}=13$ TeV. We impose a constraint
$|y|\leq y_{cut}=5$ to exclude contributions to the integral from
regions $x<10^{-5}$, which otherwise may bias the shown luminosities
at invariant masses below 40 GeV.
For $x < 10^{-5}$, the CT14 PDFs are not constrained
by the experimental data, and the final state particles are likely to be produced in the forward
region outside of the experimental acceptance of the LHC detectors.
With the constraint, the comparison of luminosities is more relevant
to the LHC measurements. We see that the Hessian and MC uncertainties
agree well across most mass range both in $gg$ and $q\bar{q}$ sectors,
with somewhat larger deviations observed at lowest and highest masses.

Taken together, Figs. \ref{fig:CT14MCasymNNLO}, \ref{fig:CT14MCsymNNLO},
and \ref{fig:CT14MCasymNLO} demonstrate that our specifications for
the construction of the replica PDFs do yield a satisfactory consistency with
the central CT14 PDFs, and with the Hessian error PDFs. Therefore
we expect that the uncertainties of cross sections calculated using the
replica PDFs will agree with those based on error propagation \textit{via}
the Hessian PDFs. Comparisons of the most common LHC cross sections and
cross section asymmetries, computed
with the CT14 Hessian and MC PDFs, are presented at \cite{CT14MCwebsite}.
By our design, in the regions where the PDFs are well constrained, the ensemble average of the predictions from either CT14MC1 or MC2 PDF set is expected to be close to the central prediction given by the central CT14 Hessian PDF set. Similarly, the asymmetric standard deviations given by the CT14MC PDF set are expected to be very close to the 68\% c.l. uncertainties of the  CT14 (Hessian) PDF error set, except when the $x$ value is very large (more than $0.7$)
or small (less than $10^{-5}$).

\begin{figure}
\includegraphics[width=0.49\textwidth]{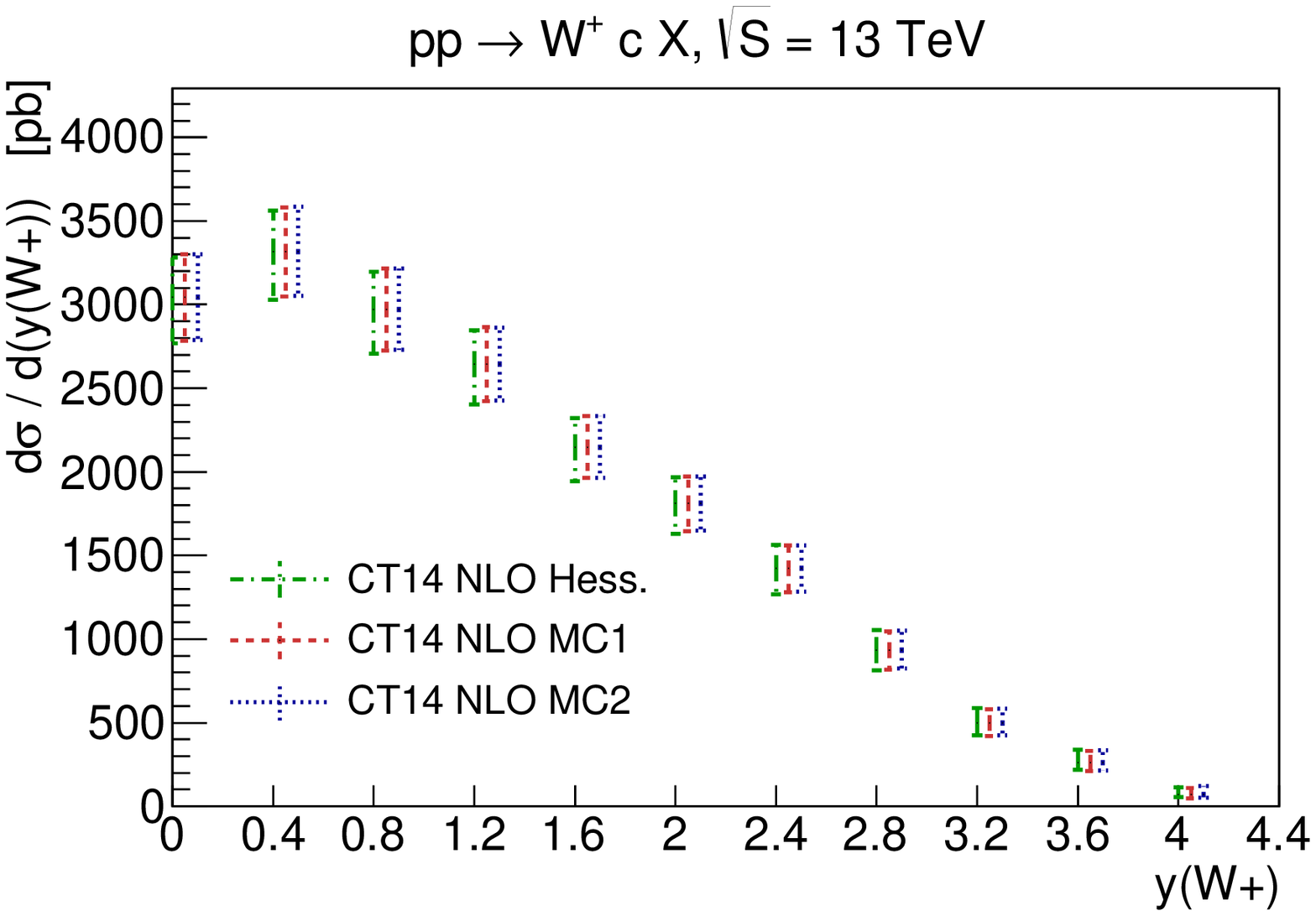}
\includegraphics[width=0.49\textwidth]{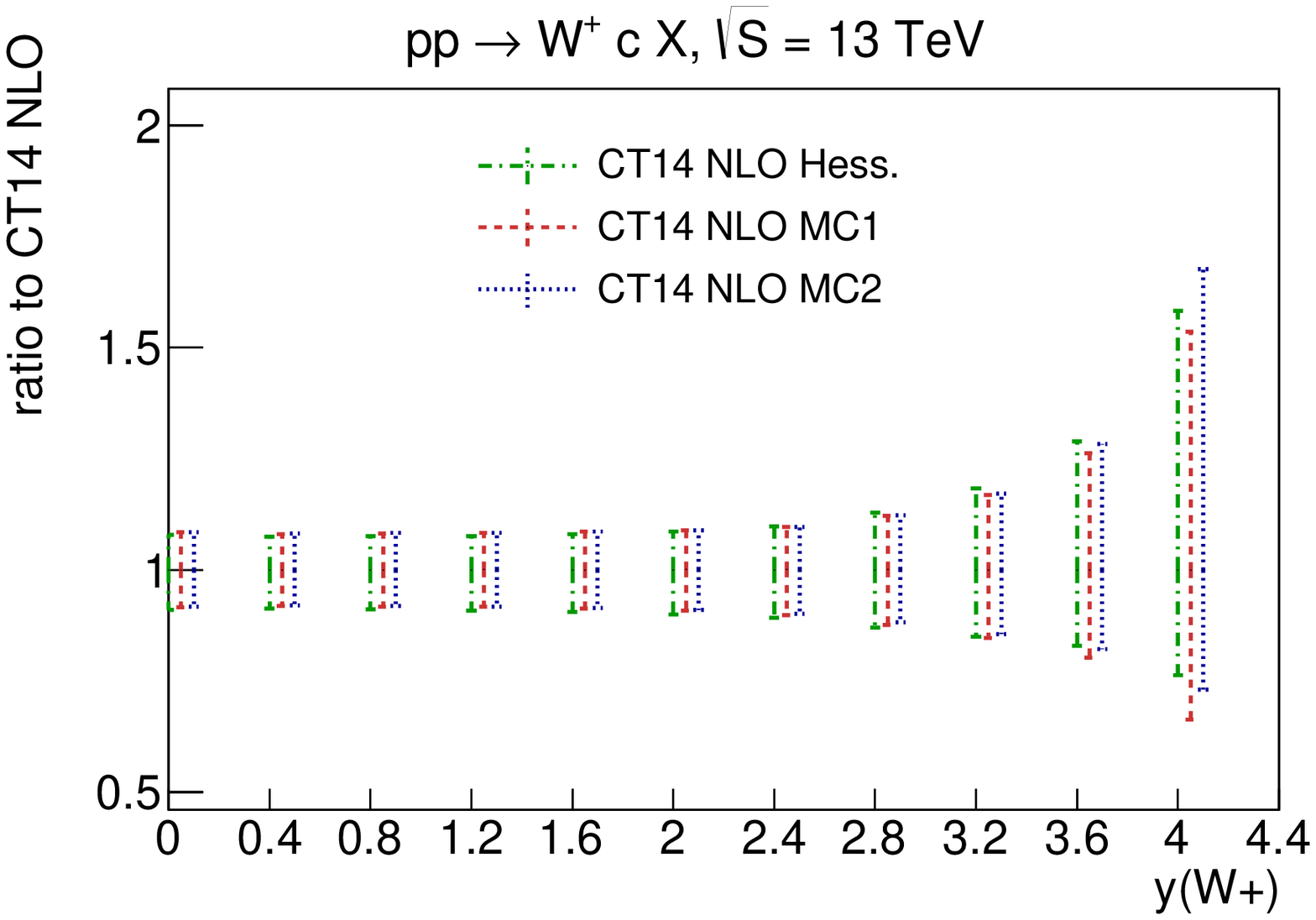}
\caption{NLO predictions for $d\sigma/dy(W^{+})$ in the process $pp \to W^{+}\bar{c}$ at the LHC 13 TeV, computed with {\sc APPLgrid}.  \label{fig:Wpc} }
\end{figure}

Some examples of cross sections are computed using the {\sc Applgrid} fast interface \cite{Carli:2010rw} for interpolation of NLO cross sections computed with the help of {\sc MadGraph\_aMC@NLO} \cite{Alwall:2014hca}, and {\sc aMCfast} \cite{Bertone:2014zva}. The {\sc Applgrid} input cross section tables are available at \cite{PDF4LHC15gallery} from a study of cross sections
based on PDF4LHC15 PDFs \cite{Badger:2016bpw,Butterworth:2015oua}.
Specifically, we computed CT14 Hessian, MC1 and MC2 rapidity distributions with no or minimal experimental cuts for production $W^{\pm}$, $Z^0$, $t\bar{t}$,
$t\bar{t} \gamma\gamma$, $W^{+}\bar{c}\, (W^{-}c)$ production at the LHC 7,8, and 13 TeV. The  
renormalization and factorization scales are 
$\mu_{R}=\mu_{F}=M_W$, $M_Z$, $H_T/2$, $H_T/2$, $M_W$, respectively. $H_T$ is the
scalar sum of transverse masses $\sqrt{p_T^2+m^2}$ of final-state
particles. For $W^{+}\bar{c}\, (W^{-}c)$ production, we neglect small
contributions with initial-state $c$ or $b$ quarks. 
For NLO single-inclusive jet and dijet production, we use public
{\sc APPLgrid} 
files in the bins of ATLAS measurements
\cite{Aad:2011fc}, created with the program {\sc NLOJET++}
\cite{Nagy:2001fj,Nagy:2003tz}. Similarly, the $W$ charge asymmetry in CMS experimental
bins \cite{Chatrchyan:2012xt,Chatrchyan:2013mza} is computed with {\sc APPLgrid} 
from \cite{Alekhin:2014irh}. An example of the cross sections on the website is shown in Fig.~\ref{fig:Wpc}. For ease of comparisons, the PDF uncertainties
are plotted both for the cross section values and
for ratios to the central CT14 prediction.

\subsection{Only an ensemble of MC replicas is meaningful \label{sec:ReplicaInterpretation}}

The results presented above provide a practical prescription for estimating
asymmetric Hessian PDF uncertainties using an ensemble of Monte-Carlo
replicas. As was already emphasized, some information about the primordial
probability distribution in the global fit is lost in the process;
however, the prescription is simple and provides a reasonable estimate
in most cases. The method is most reliable and unique when the PDF
uncertainty is small, meaning that the Taylor series converges fast.
When the uncertainty is large, we see more
variations, and several alternatives are conceivable. For example,
our MC replicas are constructed so
that their mean set coincides with the central
set of the Hessian ensemble within numerical roundoff errors, 
as long as $x$ is not too large or too small.
Instead, we could choose the
central set of the Hessian ensemble to coincide with the median
or mode sets of the MC replicas. Similarly,
the PDF uncertainty bands may be defined by the positive and negative
standard deviations discussed above, or by the boundaries of the
68\% c.l. interval centered on the ``central
PDF set'' chosen above. We have checked that the various prescriptions
agree well at intermediate $x$.

\begin{figure}
\begin{centering}
\includegraphics[width=0.46\textwidth]{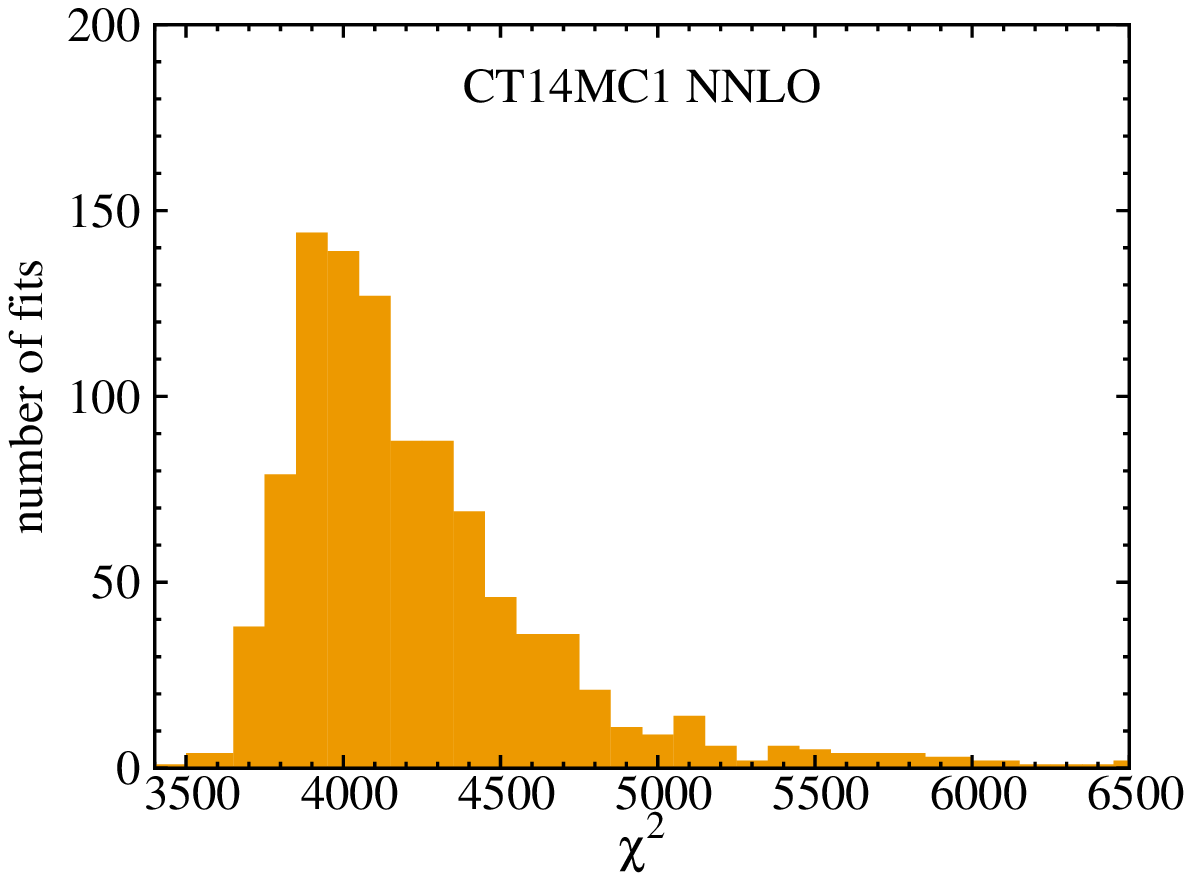}
\includegraphics[width=0.46\textwidth]{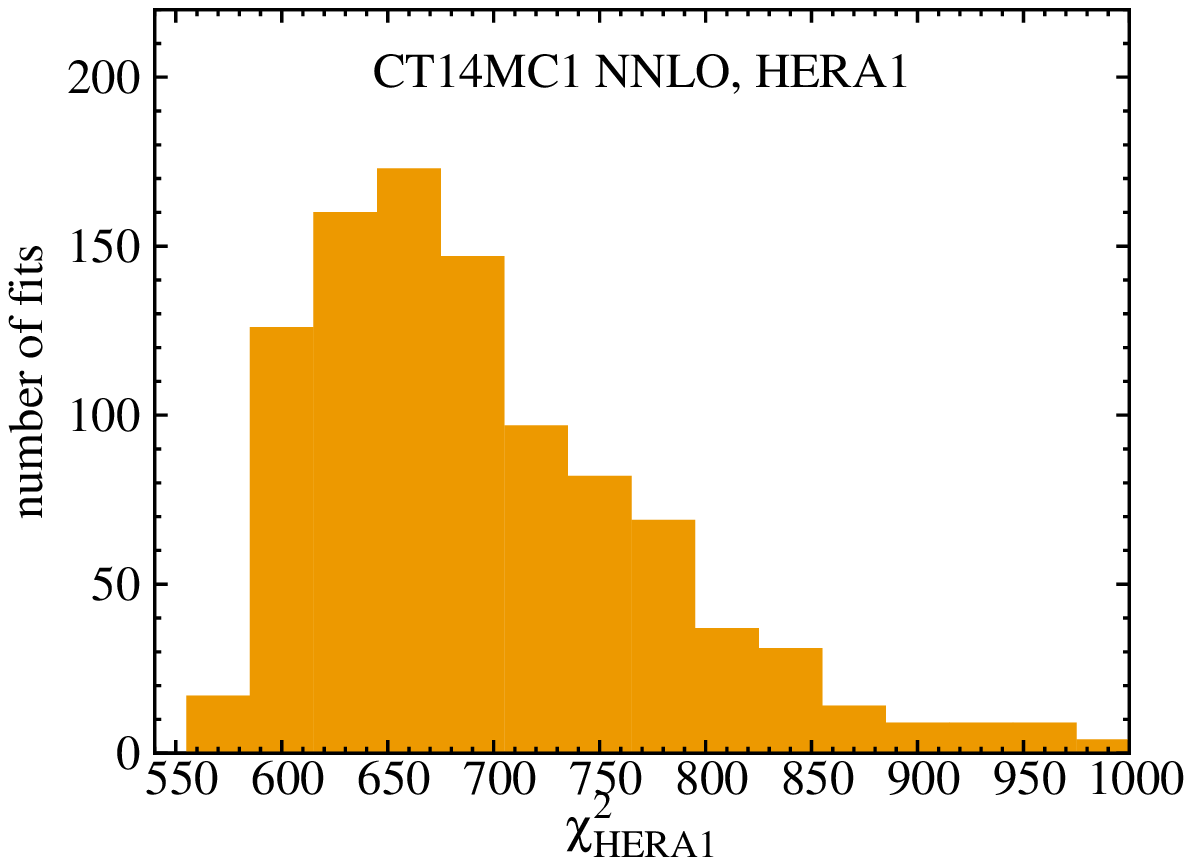}
\par\end{centering}

\begin{centering}
\includegraphics[width=0.46\textwidth]{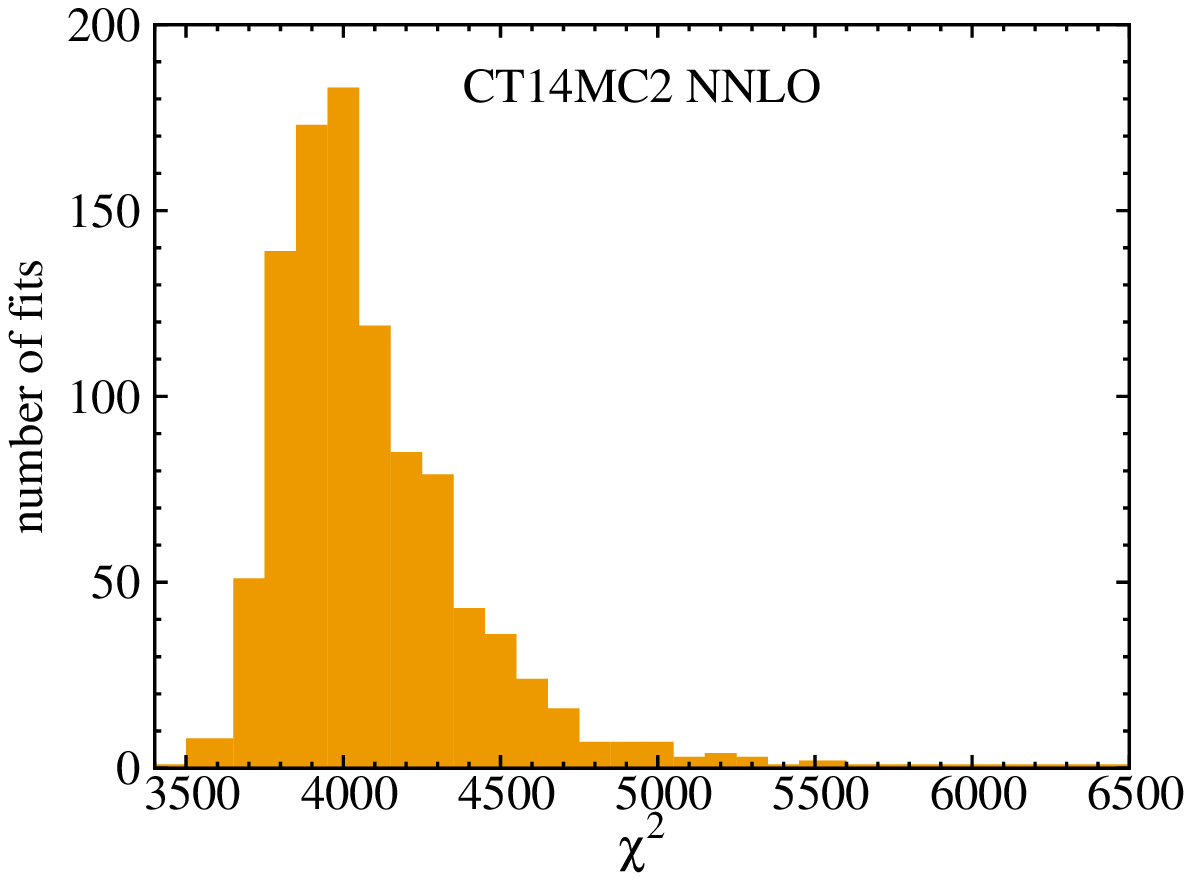}
\includegraphics[width=0.46\textwidth]{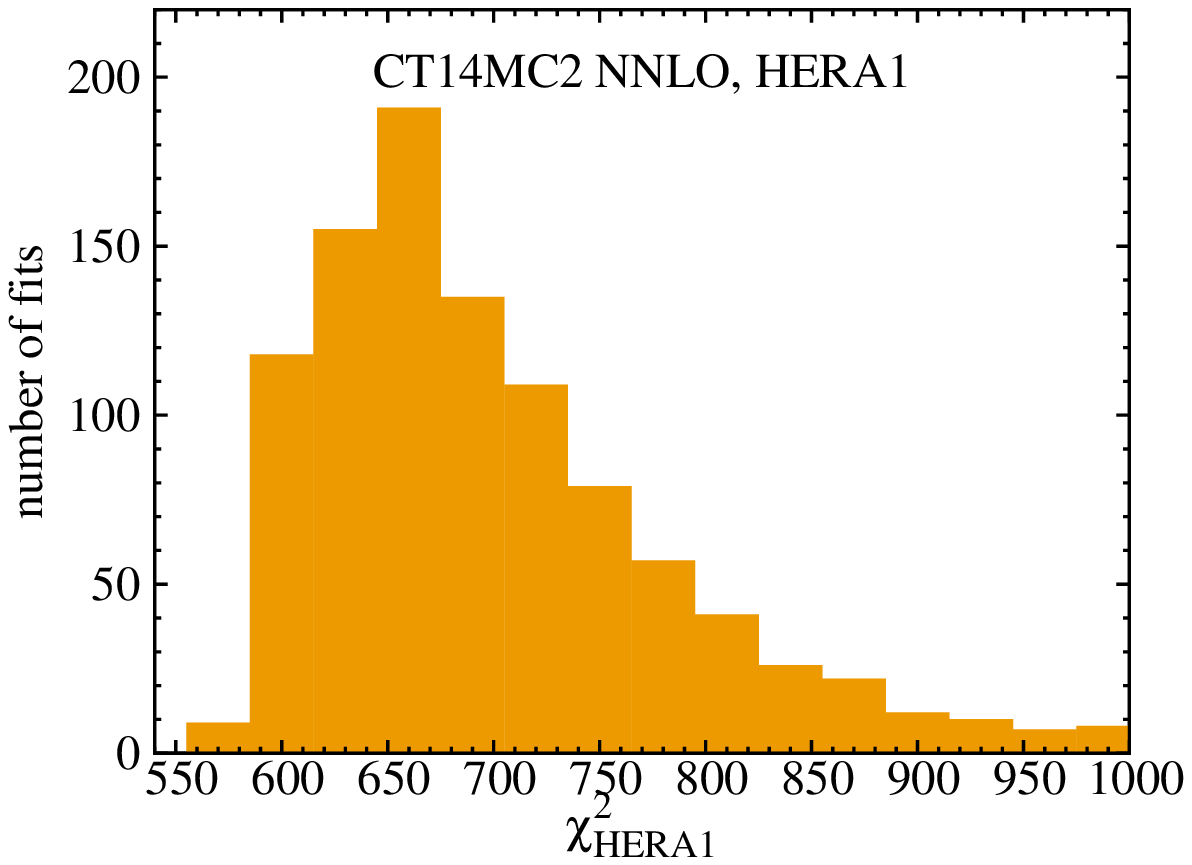}
\par\end{centering}

\caption{$\chi^{2}$ for global data (left) and combined HERA-1 data (right),
for 1000 replicas of the CT14MC NNLO ensemble. \label{fig:chi2HERA1}}
\end{figure}

It is important to note  that most replicas are poor
fits to the hadronic data used in the global analysis; however,  their averages
and standard deviations defined in Sec.~\ref{sec:sec2} provide excellent
approximations for the Hessian central PDF set and 68\% c.l. uncertainties. This
is demonstrated in Fig.~\ref{fig:chi2HERA1}, showing histograms
of $\chi^{2}$ values for the global data (3174 data points; left
panel) and for the combined HERA-1 data (579 data points \cite{Aaron:2009aa};
right panel), for the 1000 replicas in the CT14 NNLO MC1 and MC2 ensembles.
The vast majority of replicas yield very large $\chi^{2}$ values
for the global data, and even for the single experiment.
The random fluctuations of the individual replicas, which
result in large $\chi^{2}$ values for any single replica, will largely
cancel in the ensemble averages.

It is straightforward to understand why the result shown
in Fig.~\ref{fig:chi2HERA1}
occurs. Imagine that we construct a $D$-dimensional vector $\vec{a}$
whose coordinates are given by random variates $a_{i}$ sampled from
a standard normal distribution. The length of $\vec{a}$ will often
turn out to be significantly larger than 1. If those parameter values
are used for the PDFs, $\chi^{2}(\vec{a})$ will tend to be much larger
than the minimal $\chi^{2}$ in the fit, especially if the number
of dimensions $D$ is large. [Recall that the volume of an $D$-dimensional
  unit ball vanishes in the limit of large $D$. The vast majority of replicas
  will have several parameters far outside of the unit ball, i.e., away from
  the best fit by many standard deviations.]

The expected value of the length of $\vec{a}$ can be easily found as
\begin{eqnarray}
\langle|\vec{a}|\rangle & = & \frac{\int_{-\infty}^{\infty}da_{1}\int_{-\infty}^{\infty}da_{2}...\int_{-\infty}^{\infty}da_{D}\,\sqrt{\sum_{i=1}^{D}a_{i}^{2}}\,\exp\left(-\sum_{i=1}^{D}a_{i}^{2}/2\right)}{\int_{-\infty}^{\infty}da_{1}\int_{-\infty}^{\infty}da_{2}...\int_{-\infty}^{\infty}da_{D}\,\exp\left(-\sum_{i=1}^{D}a_{i}^{2}/2\right)}\nonumber \\
 & = & \left(2\pi\right)^{-D/2}\cdot\int_{0}^{\infty}da\, a^{D}\,\exp\left(-a^{2}/2\right)\cdot\int d\Omega_{D}\nonumber \\
 & = & \frac{\sqrt{2}\Gamma\left((D+1)/2\right)}{\Gamma\left(D/2\right)}\approx\sqrt{D}.\label{eq:ava}
\end{eqnarray}
The final approximation in (\ref{eq:ava}) follows from Stirling's
formula for the gamma function. For $D=28$, Eq.~(\ref{eq:ava})
gives $\langle|\vec{a}|\rangle\approx5.24$; that is, a typical displacement
vector of a CT14 replica is more than five standard units in length.
As a result, most replicas have  a $\chi^{2}$ value that significantly
exceeds the CT14 best-fit value of about 3250 for the global data
and 590 for the HERA-1 data. If the standard deviation of the normal
distribution is rather about 6 units (corresponding to $\Delta\chi^{2}\approx100$
at 90\% c.l.), the average displacement vector corresponds to $\Delta\chi^{2}\approx(5.24\cdot 6)^{2}\approx1000$
for the global data and $\Delta\chi^{2}\approx180$ (scaled down by
579/3174) for the HERA-1 data. This is far outside the typical
$1\sigma$ for the $\chi^2$ distribution, given by $T^2\approx 36$, $\sqrt{2
  N_{pts}}\approx 80 (35)$, or another common estimator!
These estimates are in  good agreement
with the actual $\chi^{2}$ averages over the CT14 replicas, denoted
as $\langle\chi^{2}\rangle_{rep},$ and equal in the case of the MC1
(MC2) NNLO ensembles to $\approx4300$ ($4200$) for the global $\chi^{2}$,
and $\approx720$ ($730$) for the HERA-1 $\chi^{2}$.

It is interesting to note \cite{ThorneDIS2016}
that the $\chi^{2}$ distribution for the
CT14 MC replicas is quite similar in shape to the $\chi^{2}$ distributions
for the global data of NNPDF replicas, obtained
with an entirely different methodology \cite{Forte:2002fg,DelDebbio:2007ee,Ball:2008by,Ball:2012cx,Ball:2014uwa}. For example, for the four NNPDF3.0
NLO fits listed in Table 1 of Ref.~\cite{Ball:2016neh}, the equivalent of 
our $\langle\chi^{2}\rangle_{rep}$ is 600-1000 units higher than
the $\chi^{2}$ value for the average PDF set of all replicas.\footnote{The NNPDF table constructs $\langle\chi^{2}\rangle_{rep}$ from the $\chi^2$ values between
  theoretical predictions for an individual PDF replica
  and true (not fluctuated) experimental data. The average is over the ensemble of the NNPDF replicas.} This
can be qualitatively understood by noticing that the tensions between
the individual experiments can be effectively accommodated by introducing
a tolerance $T_{exp}\sim2$ on the global $\chi^{2}$ at the 68\% c.l.
\cite[ sec. 7]{Pumplin:2009sc,Lai:2010vv}, and that the neural network
parametrizations effectively have of order $D=100-250$ free parameters.
This predicts $\langle\chi^{2}\rangle_{rep}-\chi_{min}^{2}\sim T_{exp}^{2}D=400-1000$,
in fair agreement with the actual NNPDF3.0 outcomes. Individual replica
PDFs thus cannot be considered as alternative PDFs with approximately
the same accuracy as the central fit, or even with the accuracy of
the Hessian error
PDFs. The replica PDFs are meaningful only inside an ensemble,
predictions based on them must be calculated by averaging over the ensemble. 

\section{Conclusions}

\label{sec:sec4}

The paper explored methods for  the conversion of Hessian PDF error sets
into Monte-Carlo replica PDFs. We observed that parameters
of the PDFs approximately obey a multi-dimensional normal distribution
in the vicinity of the global $\chi^{2}$ minimum. A statistical sample
of MC replicas dependent on the probability distribution in the $D-$dimensional
parameter space can be reconstructed under the Gaussian assumption
from the distribution of the Hessian error PDFs on the ($D-1$)-dimensional
boundary of some confidence region enclosing the best fit. The CT14
Hessian error PDFs allow the user to estimate the PDF errors
and account for the asymmetries
due to mild deviations from the Gaussian shape and nonlinear nature
of PDF parametrization. The CT14 PDFs
are parameterized to be explicitly non-negative,
as desired for obtaining sensible cross sections and uncertainties. In Sec.~\ref{sec:sec2}
we have generalized the algorithms for the generation of MC replicas and  for the estimation
of PDF uncertainties necessary to  reproduce these features. Based on these results,
we have generated two replica ensembles, designated as CT14MC1 and CT14MC2 PDFs.
The average PDF sets and the asymmetric standard deviations of MC1 and
MC2 closely reproduce the central sets and the 68\% c.l. asymmetric uncertainties
of the respective Hessian ensembles, cf. Sec.~\ref{sec:sec3}. With the
naive conversion, the mean set over the replicas may
deviate from the central Hessian PDF set because of the truncation of
Taylor series. After we apply a correction,
the MC mean and central Hessian sets coincide, preventing
potential discrepancies in the predictions. 
In addition, the CT14MC2 PDF replicas, obtained by sampling
of a log-normal distribution, are explicitly non-negative.

To achieve good agreement between the CT14 Hessian and MC uncertainty bands,
we recommend to construct the MC replicas by random sampling according to the algorithm summarized as Eqs.~(\ref{Xk}, \ref{Delta}, \ref{dXkAsym}),
and Sec.~\ref{sec:MCPDFs}. The central PDF set of such MC ensemble by construction coincides with the central Hessian PDF set. The PDF uncertainties for the MC ensemble are to be estimated by asymmetric standard deviations (\ref{dMCXp}, \ref{dMCXm}), introduced as
the counterparts for the master formulas (\ref{dH68Xp}, \ref{dH68Xm})
for the Hessian asymmetric errors. 

Our results in Sec.~\ref{sec:sec2} 
clarify several aspects of the replica generation
method that were not addressed in the previous work.  
Section~\ref{sec:sec3} demonstrates that, in most of the $x$ range, where
the PDF uncertainties are small, we observe good agreement between
the CT14 Hessian, MC1, and MC2 error bands. In the extrapolation regions,
where the linear approximations cease to be adequate, differences
between the Hessian, MC1 and MC2 error bands are indicative of intrinsic
ambiguities in the replica generation methods. At such extreme $x$,
they may affect the combination
of the Hessian and MC ensembles, as in the recommended PDF4LHC procedure,
or PDF replica re-weighting. These ambiguities require consideration
when the converted replicas are utilized in PDF reweighting or PDF
combinations. The positivity
constraint on CT14 MC2 yields more physical behavior in poorly
constrained $x$ intervals.

Once the MC replicas are generated by conversion, we
examine their statistical properties in
Sec.~\ref{sec:ReplicaInterpretation}.
We point out that many individual MC replica sets yield poor $\chi^2$, so
only their statistical combinations, such as the mean, standard
deviation, etc., are meaningful in applications. [The majority of MC
  replicas aren't good fits, their combination is.]
On the other hand, the $\chi^2$
distributions of MC replicas obtained by conversion (such as CT14 MC) or genetic
algorithm (such as NNPDF) are very similar. This reflects 
the underlying commonalities of the two methods, leading to
comparable PDF errors in the two approaches
in spite of the distinct error definitions and fitting procedures.

The CT14 MC1 and MC2 ensembles at NNLO and NLO accuracy, together
with a fast standalone driver program for their interpolation, can
be downloaded at \cite{CT14website}. They are also distributed as
a component of the LHAPDF6 library \cite{LHAPDFwebsite}. A public C++
code \textsc{mcgen }is made available \cite{MCGENwebsite} for generation
of MC PDF replicas using the normal, log-normal, and Watt-Thorne sampling
methods, and with or without including the $\Delta$ shifts. \textsc{mcgen}
can be run as a standalone program or together with the Mathematica
package MP4LHC for combination of PDF ensembles according to the meta-parametrization
method \cite{Gao:2013bia}. After $N_{input}$ Hessian error PDF sets
are read in the form of LHAPDF6 grids, $N_{rep}$ output replicas
are generated by random displacements of the Hessian replicas. Besides
the replica generation, \textsc{mcgen} supports various algebraic
operations with PDFs in the format of LHAPDF6 grid files, such
as addition, averaging, and multiplication of the tables in which
the PDF values are stored.

\subsection*{Acknowledgments}

We thank R. Thorne for an instructive discussion about statistical
properties of MC replicas.
J.H., J.G., and P.N. thank the Kavli Institute for Theoretical Physics at Santa Barbara, CA and organizers of the ``LHC Run II and Precision Frontier'' research program for hospitality and productive atmosphere during completion of this paper. This research was supported in part by the National Science Foundation
under Grants No. PHY-1410972 and PHY11-25915; by the
U.S. Department of Energy under Contract No. DE-AC02-06CH11357 and Grants
DE-SC0013681 and DE-SC0010129; and by the
 National Natural Science Foundation of China under Grant No. 11465018.

 \appendix
 \section*{Definition of $\chi^2$ in CT fits}
 For completeness of the presentation, in this appendix,
 we reproduce the definition for the figure-of-merit
 function $\chi^2$ and the procedure for determination of PDF
 uncertainties in the CT14 fit, introduced in full detail in 
 Refs.~\cite{Lai:2010vv,Gao:2013xoa,Dulat:2015mca}.

The most probable solutions for CT14 PDFs are found by minimization
of a \emph{global} log-likelihood function 
\begin{equation}
\chi_{global}^{2}=\sum_{n=1}^{N_{exp}}\chi_{n}^{2}+\chi_{th}^{2},\label{Chi2global}
\end{equation}
which sums contributions $\chi_{n}^{2}$ from $N_{exp}$ fitted experiments,
and a contribution $\chi_{th}^{2}$ specifying theoretical
conditions (``Lagrange Multiplier constraints'') imposed on some
PDF parameters. In turn, the terms $\chi_{n}^{2}$ are given by 
\begin{equation}
\chi_{n}^{2}(\vec{a},\vec{\lambda})=\chi_{D}^{2}+\chi_{\lambda}^{2},\label{Chi2n}
\end{equation}
where 
\begin{equation}
\chi_{D}^{2}\equiv\sum_{k=1}^{N_{pts}}\frac{1}{s_{k}^{2}}\left(D_{k}-T_{k}(\vec{a})-\sum_{\alpha=1}^{N_{\lambda}}\beta_{k,\alpha}\lambda_{\alpha}\right)^{2},\label{Chi2D}
\end{equation}
and 
\begin{equation}
\chi_{\lambda}^{2}\equiv\sum_{\alpha=1}^{N_{\lambda}}\lambda_{\alpha}^{2}.\label{Chi2lambda}
\end{equation}
The $\chi_{n}^{2}$ contribution is a function of the PDF parameters
$\vec{a}$ and systematic nuisance parameters $\vec{\lambda}$. For a
$k$-th data point, $T_{k}$, $D_{k},$ $s_{k,}$ and $\beta_{k\alpha}$
are the theoretical prediction, central experimental value, uncorrelated
experimental uncertainty, and systematic correlation matrix, respectively. 

The minimum of the $\chi_{global}^{2}$ function is found iteratively
by the method of steepest descent using the program \texttt{MINUIT}.
The boundaries of the 90\% c.l. region around the minimum of $\chi_{global}^{2}$,
and the eigenvector PDF sets quantifying the associated uncertainty,
are found by iterative diagonalization of the Hessian matrix \cite{Stump:2001gu,Pumplin:2001ct},
which finds independent, or eigenvector, combinations of the PDF parameters
$a_{i}$. These combinations are denoted by $z_{i}$, with the best-fit
parameter combination corresponding to $z_{1}=z_{2}=...=0.$ The 90\%
c.l. boundary around the best fit is determined by applying two tiers
of criteria, based on the increase in the global $\chi_{global}^{2}$
summed over all experiments, and the agreement with individual experimental
data sets \cite{Lai:2010vv,Gao:2013xoa,Dulat:2013hea}. The first
type of condition demands that the global $\chi^{2}$ does not increase
above the best-fit value by more than $\Delta\chi^{2}=T^{2}$, where
the 90\% C. L. region corresponds to $T\approx10$. The second condition
introduces a penalty term $P$ in $\chi^{2}$ when establishing the
confidence region, which quickly grows when the fit ceases to agree
with any specific experiment within the 90\% c.l. for that experiment.
The effective function $\chi_{\mathrm{e}ff}^{2}=\chi_{global}^{2}+P$
is scanned along each eigenvector direction until $\chi_{eff}^{2}$
increases above the tolerance bound, or rapid $\chi_{eff}^{2}$ growth
due to the penalty $P$ is triggered.

The penalty term is constructed as 
\begin{equation}
P=\sum_{n=1}^{N_{exp}}(S_{n})^{p}\theta(S_{n})
\end{equation}
from the equivalent Gaussian variables $S_{n}$ that obey an approximate
standard normal distribution independently of the number of data points
$N_{ptx,n}$ in the experiment. Every $S_{n}$ is a monotonically increasing
function of the respective $\chi_{n}^{2}$ given in \cite{Dulat:2013hea,Lewis:1988}.
The power $p=16$ is chosen so that $(S_{n})^{p}$ sharply increases
from zero when $S_{n}$ approaches 1.3, the value corresponding to
90\% c.l. cutoff.

\newpage

\end{document}